\documentclass[sn-mathphys,Numbered]{sn-jnl}
\usepackage{graphicx}%
\usepackage{multirow}%
\usepackage{amsmath,amssymb,amsfonts}%
\usepackage{amsthm}%
\usepackage{mathrsfs}
\usepackage[title]{appendix}%
\usepackage{xcolor}%
\usepackage{textcomp}%
\usepackage{manyfoot}%
\usepackage{booktabs}%
\usepackage{algorithm}%
\usepackage{algorithmicx}%
\usepackage{algpseudocode}%
\usepackage{listings}%
\usepackage{subfig}

\theoremstyle{remark}
\newtheorem{example}{Example}
\renewcommand{\theexample}{(\alph{example})}

\DeclareMathOperator\arctanh{arctanh}
\DeclareSymbolFontAlphabet{\mathrsfs}{rsfs}
\DeclareMathAlphabet{\mathcal}{OMS}{cmsy}{m}{n}
\newcommand{\scri}{\mathrsfs{I}} 
 \newcommand{\be}{\begin{equation}}
\newcommand{\ee}{\end{equation}}

\newcommand{\gp}{{\textrm{GP}}}
\newcommand{\ef}{{\textrm{EF}}}

\newcommand{\widebar}[1]{\mkern 2mu\overline{\mkern -2mu#1\mkern -2mu}\mkern 2mu}

\newtheorem{definition}{Definition}

\begin{document}

\title[Regularization of null horizons]{Bridging time across null horizons}

\author{\fnm{An\i l} \sur{Zengino\u{g}lu}}\email{anil@umd.edu}

\affil{\orgdiv{Institute for Physical Science and Technology}, \orgname{University of Maryland}, \orgaddress{\city{College Park}, \postcode{20742}, \state{MD}, \country{USA}}}

\abstract{

General relativity, as a diffeomorphism-invariant theory, allows the description of physical phenomena in a wide variety of coordinate systems. In the presence of boundaries, such as event horizons and null infinity, time coordinates must be carefully adapted to the global causal structure of spacetime to ensure a computationally efficient description. Horizon-penetrating time is used to describe the dynamics of infalling matter and radiation across the event horizon, while hyperboloidal time is used to study the propagation of radiation toward the idealized observer at null infinity. \\

In this paper, we explore the historical and mathematical connection between horizon-penetrating and hyperboloidal time coordinates, arguing that both classes of coordinates are simply regular choices of time across null horizons. We review the height-function formalism in stationary spacetimes, providing examples that may be useful in computations, such as source-adapted foliations or Fefferman-Graham-Bondi coordinates near null infinity. We discuss bridges connecting the boundaries of spacetime through a time hypersurface across null horizons, including the event horizon, null infinity, and the cosmological horizon. \\

This work is motivated by the broader effort to understand the role of time in general relativity and reviews a unified framework for handling null boundaries in analytical and numerical approaches. The insights developed here offer practical tools for numerical relativity, gravitational wave astronomy, and other explorations of the large-scale structure of spacetimes.
}


\maketitle

\pagebreak
\tableofcontents
\pagebreak
\section{Introduction}\label{sec:intro}

What is time? This work is motivated by the broader effort to address this timeless question. Though the problem of time may seem rather philosophical with little practical relevance, recent explorations of this question within black-hole physics have led to powerful computational tools to study wave propagation phenomena. 


Time, in general relativity, is described by a coordinate whose level sets are spacelike hypersurfaces, but is otherwise arbitrary, in accordance with general covariance.  A natural, partial choice for time exists in perturbation theory where we study dynamic perturbations of a stationary background: we adapt the \textit{passage} of time to the timelike Killing field. However, there is still a large freedom in the choice of \textit{moments} of time, which amounts to the choice of distant simultaneity between spacelike separated events. In this paper, we review a proposed family of time hypersurfaces for black-hole perturbation theory \cite{zenginoglu_hyperboloidal_2008, Zengino_lu_2011, macedo2020hyperboloidal, PanossoMacedo:2023qzp}.

It is often argued that choices of coordinates are merely a matter of convenience, irrelevant for physical observables. The observable statements of the theory are invariant under smooth, invertible coordinate transformations. 
However, the choice of suitable \textit{families} of coordinates may have physically relevant consequences beyond mere convenience or computational efficiency. Physical observables in different classes of coordinate systems related to each other by \textit{singular} transformations may describe physically distinct behaviors. As a prominent example, consider global energy in asymptotically flat spacetimes. Energy at spatial infinity is conserved in time but energy evolving along null infinity decays \cite{ashtekar2024operational}. This is a difference in the description of a physically observable quantity arising from distinct boundary structures that break general covariance. If the coordinates are not adapted to these structures, the physically relevant behavior must be imposed \textit{by hand}. Instead, using boundary-adapted coordinates provides a natural, geometric description of the physically expected behavior without artificial conditions. In this paper, we discuss the ideas underlying families of boundary-adapted time coordinates for the simplest case of stationary and spherically symmetric spacetimes in the presence of horizons.

A horizon is a limiting surface of no return—a boundary beyond which classical interaction is impossible. We can send signals to the horizon but can never receive signals back, preventing the usual two-way synchronization of time. The Schwarzschild event horizon was the first relativistic horizon to be identified, and it caused much confusion in the early days of relativity \cite{israel1987dark, senovilla_1965_2015, nielsen2022origin, landsman_penroses_2022}. The nature of the apparent Schwarzschild singularity at the event horizon was not understood for many decades. Today, we recognize this singularity as a coordinate failure due to the intersection of Schwarzschild time slices at the bifurcation sphere—an artifact of time symmetry. In realistic scenarios, such as black holes formed by stellar collapse, the time-reflection symmetry is broken and no bifurcation sphere exists. We remove the coordinate singularity by adopting horizon-penetrating coordinates that bypass the bifurcation sphere and do not impose time-reflection symmetry. In Sec.~\ref{sec:2}, we review the developments that clarified the global causal structure of Schwarzschild spacetime.

Penrose's groundbreaking work \cite{penrose_asymptotic_1963,penrose_zero_1965} revealed that asymptotic null boundaries exist even in the flat spacetime of special relativity. His conformal extension of flat spacetime demonstrated that standard Cauchy time slices become singular at spatial infinity\footnote{We refer here to the intersection of standard time slices at spatial infinity, which is a coordinate singularity present even in flat spacetime. The general singularity structure of asymptotically flat solutions to the Einstein equations at infinity is a deep and subtle problem that goes beyond coordinates \cite{friedrich1998gravitational, kehrberger2021case}.}—a singularity that can be regularized using Penrose time slices\footnote{Penrose time is the time coordinate used to draw conformal diagrams (see Sec.~\ref{sec:penrose}).}. Although Penrose's conformal method is now a standard tool for understanding global structure \cite{kroon2017conformal}, the associated coordinates are rarely used in practical computations because they are not adapted to the timelike Killing field of stationary backgrounds. In Sec.~\ref{sec:3}, we review the construction of regular, \textit{stationary} time coordinates across null infinity in Minkowski spacetime.

The two null boundaries discussed in Secs.~\ref{sec:2} and \ref{sec:3}—black-hole horizon and null infinity—share causal and geometric properties. In Sec.~\ref{sec:4}, we demonstrate that horizon-penetrating time is conceptually analogous to hyperboloidal time in that the level sets are non-intersecting and transverse across null horizons. The primary advantage of such coordinates lies in their regularity, making them useful for theoretical analysis and practical computation. Horizon-penetrating coordinates can be viewed as hyperboloidal compactification, placing them within the broader historical context of resolving coordinate singularities, and suggests new approaches for developing coordinate systems suitable for applications. The unified framework of null-transverse hypersurfaces extends naturally to cosmological spacetimes. While such hypersurfaces are often called hyperboloidal across the cosmological horizon, they should be more appropriately referred to as horizon-penetrating because the cosmological horizon is at a finite proper distance and the associated null-transverse hypersurfaces do not have asymptotically hyperbolic geometry. In Sec.~\ref{sec:4}, we discuss horizon-penetrating coordinates across the cosmological horizon in de Sitter spacetime.

In Sec.~\ref{sec:5}, we combine the ideas from the previous sections in the notion of a bridge that connects null boundaries via non-intersecting, null-transverse hypersurfaces. Bridges link the black-hole horizon and the observer, reflecting correlations between these null horizons in the solutions to the field equations. Causal similarities between finite horizons in physical space and infinite horizons in conformal space also include the cosmological horizon for non-vanishing cosmological constant and have been a topic of recent interest \cite{ashtekar2024horizons, ashtekar2024null,freidel2024renormalization}. For example, the black hole tomography programme investigates correlations between the gravitational wave signal observed at null infinity and the infalling radiation at the horizon \cite{Ashtekar_2022, metidieri2025blackholetomographyunveiling} based on the weakly isolated horizon framework \cite{ashtekar2004isolated, ashtekar2024wih}. In Sec.~\ref{sec:5}, we review examples of stationary bridge foliations in Schwarzschild and Schwarzschild--de Sitter spacetimes.

The goals of this paper are as follows. First, we place recent developments in the hyperboloidal method within a broader historical context, highlighting the parallels between horizon penetration and hyperboloidal compactification. 
Second, we review different classes of regular, null-transverse foliations common to both horizon-penetrating and hyperboloidal coordinates, such as spatially flat and characteristic-preserving foliations. Third, we develop the concept of bridges to capture the essential features of a regular time coordinate in the presence of null boundaries, both for isolated systems and for cosmological spacetimes. Finally, we introduce new examples of time coordinates that may be useful for future applications, including source-adapted  bridge foliations and Fefferman-Graham-Bondi-type coordinates.





\section{Singularity resolution at the Schwarzschild radius}\label{sec:2}

In this section, we review the resolution of the coordinate singularity at the bifurcation sphere and the global causal structure of Schwarzschild spacetime to put horizon-penetrating and hyperboloidal time into historical and mathematical context.

\subsection{The historical timeline}

Following Schwarzschild's publication of the spherically symmetric vacuum solution of the Einstein equations in 1916, the coordinate singularity at the event horizon created much confusion in general relativity \cite{israel1987dark, nielsen2022origin, landsman_penroses_2022, senovilla_1965_2015}. The metric describing Schwarzschild spacetime is commonly written in Schwarzschild-Droste coordinates\footnote{Schwarzschild discovered the solution while at the Eastern Front in 1915, using a different radial coordinate. The metric now known as the Schwarzschild metric was written by Droste, a student of Lorentz, who independently derived the solution. He also provided a detailed analysis of its timelike and null geodesics, identified the photon sphere, and gave an operational definition for the familiar areal coordinate $r$, which we use today \cite{droste1917field}.},
\begin{equation} \label{eqn:schwarzschild}
    ds^2 = - f(r) dt^2 + \frac{1}{f(r)} dr^2  + r^2 d\omega^2, \quad \textrm{with} \quad f(r) = 1-\frac{2M}{r},
\end{equation}
where $d\omega^2 = d\theta^2 + \sin^2\theta\, d\varphi^2$ is the standard metric on the unit sphere. The source of the confusion was the singularity of this metric at the Schwarzschild radius, $r_b = 2M$.

A regular expression across the Schwarzschild radius was discovered independently by Gullstrand \cite{gullstrand1922allgemeine} and Painlevé \cite{painleve1922mecanique} in 1921. Neither Gullstrand nor Painlevé recognized that their metric described the Schwarzschild spacetime in different coordinates\footnote{This confusion over coordinates is why Einstein did not receive a Nobel Prize for relativity. Gullstrand, who was on the Nobel committee, opposed the confirmation of the general relativistic prediction of Mercury's perihelion advance because of his misunderstanding of coordinate freedom \cite{nielsen2022origin}.}. Shortly after, in 1924, Eddington discovered another regular expression, though he also did not make the connection \cite{eddington1924comparison}. In 1933, Lemaître demonstrated that the singularity at the Schwarzschild radius was fictitious and could be removed by a suitable choice of coordinates \cite{lemaitre1933univers, lemaitre_expanding_1997}. He noted that the coordinate singularity at the black hole horizon was similar to the singularity at the cosmological horizon. This similarity will be a recurring theme of this paper.

Despite Lemaître's insights, confusion over horizon singularities persisted until the 1950s, when the complete geometry of Schwarzschild spacetime was understood through the works of Synge \cite{synge_gravitational_1950}, Finkelstein \cite{finkelstein_past-future_1958}, Fronsdal \cite{fronsdal_completion_1959}, Kruskal \cite{kruskal_maximal_1960}, and Szekeres \cite{szekeres1960singularities, szekeres_golden_2002}. Our modern picture for the global causal structure of spacetimes, essential for understanding null horizons, was finally revealed by Penrose in the mid-1960s \cite{penrose_asymptotic_1963, penrose_zero_1965}.

A central thesis of this paper is that resolving the coordinate singularity at the Schwarzschild radius has a similar mathematical structure to resolving the coordinate singularity at spatial infinity. Therefore, it is instructive to recap the construction of regular time coordinates across the Schwarzschild event horizon. We will generalize elements of this procedure in later sections to null-transverse coordinates.

\subsection{Early regularizations preserving time-translation symmetry}\label{sec:early_developments}
As Lemaître noted in 1933 \cite{lemaitre1933univers, lemaitre_expanding_1997}, the coordinate singularity at the Schwarzschild radius arises from the requirement for a \textit{static} time coordinate. Schwarzschild originally sought a spherically symmetric, time-independent solution to the Einstein equations, with coordinates adapted to these symmetries. The requirement of staticity implies time-reversal symmetry, leading to a vanishing diagonal space-time component. By relaxing this requirement, the coordinate singularity at the event horizon can be removed while still preserving the time-independence of the metric in the exterior Schwarzschild region ($r>2M$). As a first example, consider the Gullstrand--Painlevé (GP) time coordinate. GP time is adapted to infalling particles, measures proper time along radial geodesics into the black hole, and remains regular across the horizon \cite{martel2001regular}. The resulting GP metric\footnote{The expression ``GP metric" is shorthand for the Schwarzschild metric in GP coordinates.} is time-independent and includes a non-diagonal term,
\begin{equation}\label{eqn:gp} 
    ds^2_{\gp} =  -f dt_\gp^2 + 2 \sqrt{\frac{2M}{r}} dt_\gp dr + dr^2 + r^2 d\omega^2. 
\end{equation}
The geometry of spatial slices in GP time is intrinsically flat, reflecting that a freely falling particle experiences no gravitational force—Einstein's ``happiest thought'' \cite{worden2022einstein}. We refer to time transformations satisfying this property as \textit{spatially flat}.

The GP metric shows that the event horizon is a regular hypersurface in Schwarzschild spacetime for freely falling observers. It is related to the Schwarzschild metric \eqref{eqn:schwarzschild} through the time transformation,
\begin{equation}\label{eqn:gp_trafo} 
    t_{\gp} = t + \int \frac{\sqrt{1-f}}{f} dr = t_\textrm{S} + 2 \sqrt{2Mr} + 2 M \ln\left|\frac{\sqrt{\frac{r}{2M}}-1}{\sqrt{\frac{r}{2M}}+1}\right|. 
\end{equation}
This transformation is logarithmically singular at $r_b=2M$ as $\mathcal{O}(\ln |r-r_b|)$. This singularity is due to the nature of Schwarzschild time. A singular transformation is required to obtain a ``regular" coordinate from a ``singular" one\footnote{The terms ``regular" and ``singular" refer to Synge's methodological criticism \cite{synge_model_1974}, which can be traced back to Hilbert's incorrect argument that the transformation to resolve the singularity should be regular \cite{landsman_penroses_2022}.}.

We next discuss the representation of the Schwarzschild metric, initially given by Eddington in 1924 \cite{eddington1924comparison}, but fully appreciated only after Finkelstein's rediscovery in 1958 \cite{finkelstein_past-future_1958}. The Schwarzschild metric in Eddington--Finkelstein (EF) time reads,
\begin{equation} \label{eqn:ef} 
    ds^2_{\ef} =  -f dt_\ef^2 + \frac{4M}{r} dt_\ef dr + \left(1+\frac{2M}{r}\right) dr^2 + r^2 d\omega^2.
\end{equation}
To understand the relationship between the EF and Schwarzschild time coordinates, we introduce the tortoise coordinate,
\begin{equation}\label{eqn:tortoise} 
    r_\ast = \int \frac{dr}{f} = r + \frac{1}{2\kappa_b} \ln\left| \frac{r}{r_b}-1 \right|, \qquad \kappa_b := \frac{1}{2} \frac{df}{dr}\Big|_{r=r_b} = \frac{1}{4M},
\end{equation}
where $\kappa_b$ is the surface gravity and the integration constant has been chosen so that $r_\ast(0) = 0$. The tortoise coordinate effectively captures the behavior of radial null surfaces. Outgoing null surfaces are given by $u=t-r_\ast$; ingoing null surfaces are given by $v=t+r_\ast$. EF time is adapted to ingoing null rays instead of infalling particles:
\begin{equation}\label{eqn:ef_trafo} 
    t_\ef  + r = t + r_\ast \quad \implies \quad t_\ef = t + \frac{1}{2\kappa_b} \ln\left|\frac{r}{r_b}-1\right|.
\end{equation}
In other words, the form of the ingoing null rays in terms of EF time and areal coordinate is the same as the form of the ingoing null rays in terms of Schwarzschild time and tortoise coordinate. This implies that the ingoing characteristic speed remains the same in both coordinate systems. We refer to time transformations satisfying this property as \textit{characteristic-preserving}. Note that the tortoise coordinate $r_\ast$ is unbounded toward the event horizon where the areal coordinate $r$ has a finite value.  

\begin{figure}[h]%
    \vspace{-5mm}
    \includegraphics[width=0.33\textwidth]{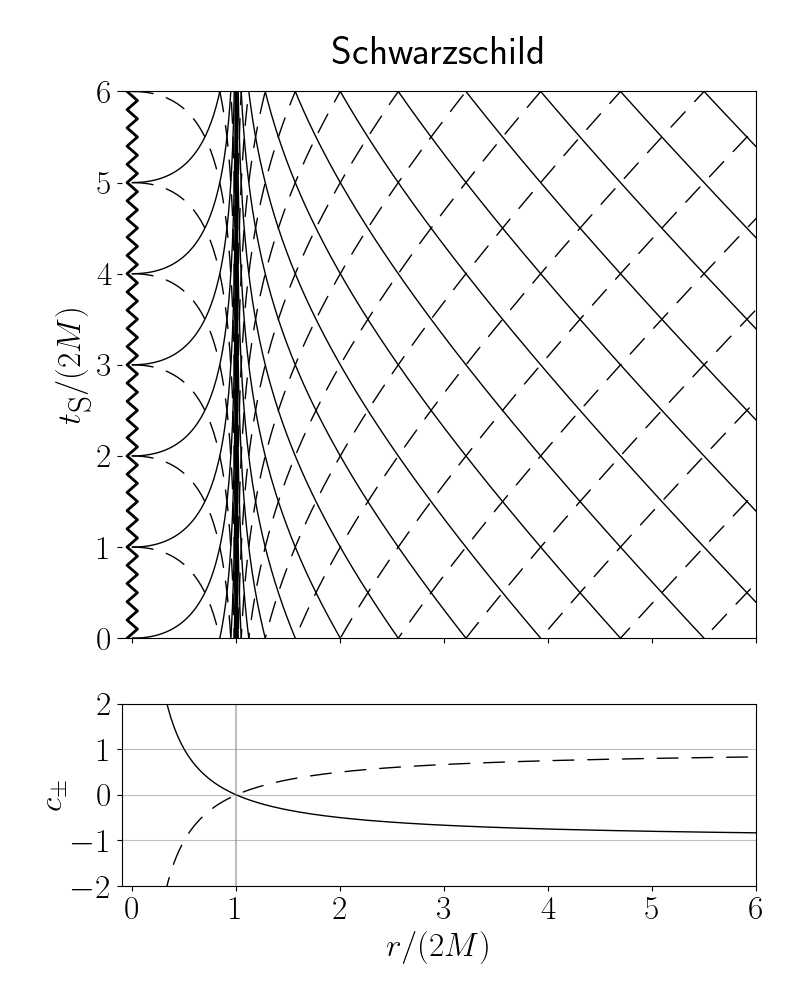}\hspace{-2mm}
    \includegraphics[width=0.33\textwidth]{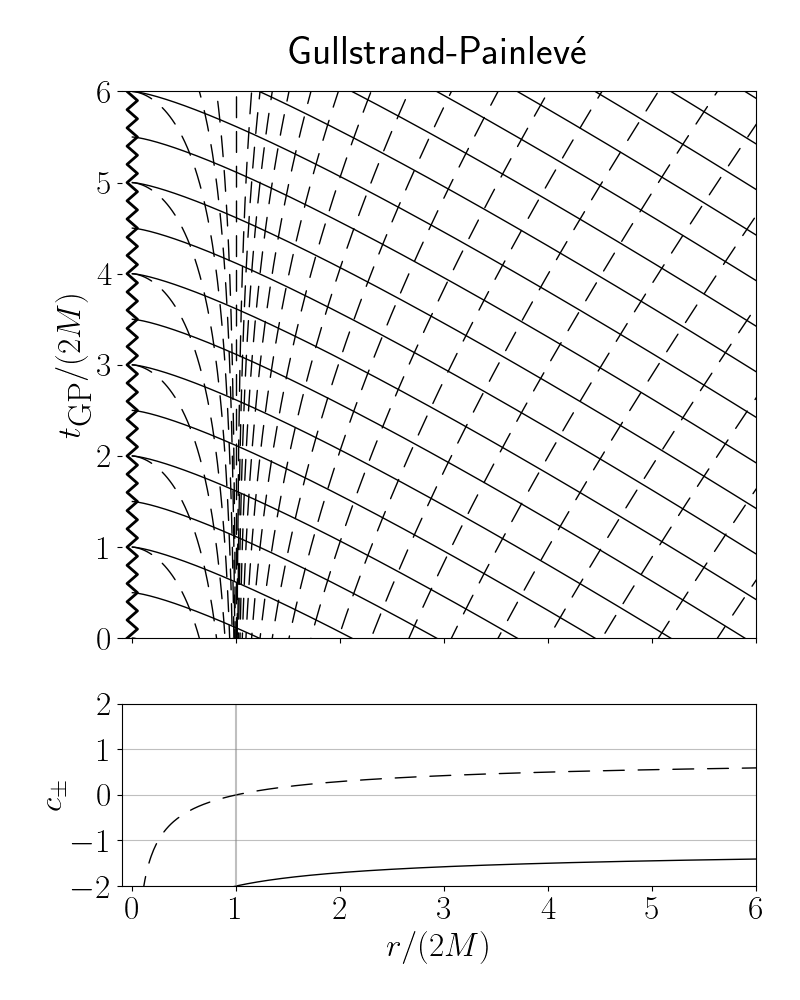}\hspace{-2mm}
    \includegraphics[width=0.33\textwidth]{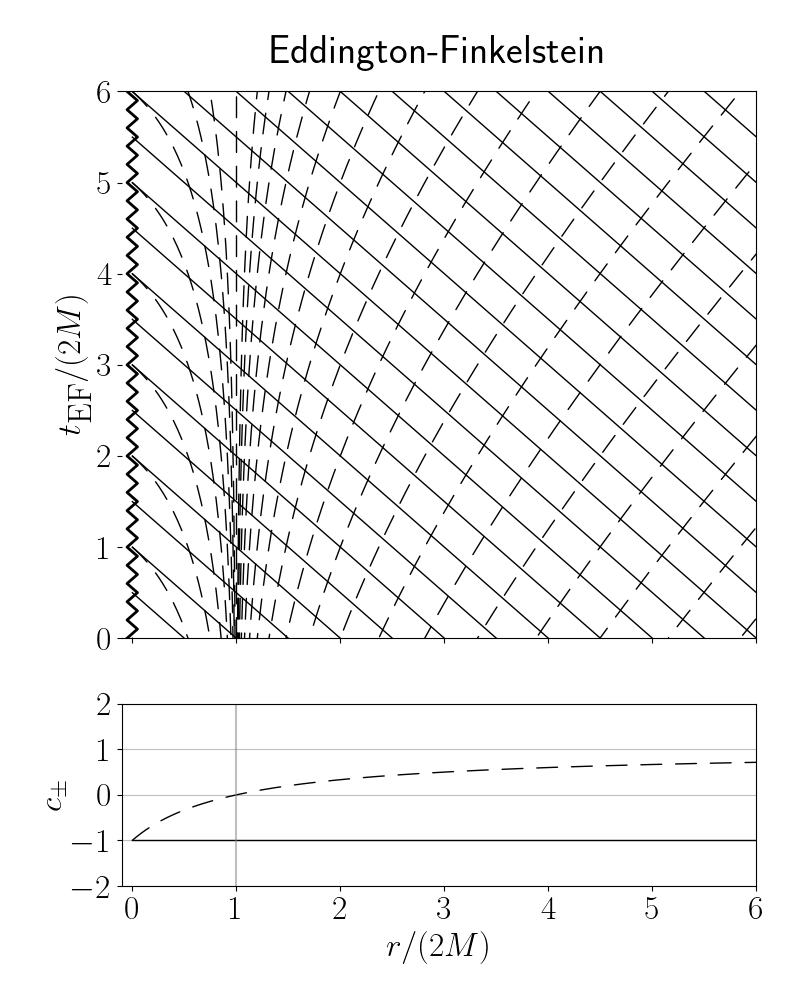}
    \caption{Characteristics in Schwarzschild (left), Gullstrand--Painlevé (middle), and Eddington--Finkelstein (right) coordinates in units of Schwarzschild radius. There are no outgoing characteristics (dashed lines) across the Schwarzschild horizon in accordance with the intuition that no light can escape a black hole. However, for Schwarzschild time, there are also \textit{no ingoing characteristics} (solid lines) across the horizon, in conflict with our understanding of a black hole. Ingoing characteristics enter the black hole region smoothly in EF and GP time. The right figure shows that the ingoing characteristic speed is one in EF time.}
    \label{fig:chars}
    \vspace{-7mm}
\end{figure}

Figure \ref{fig:chars} illustrates the properties of radial null rays in the three time-independent coordinate systems discussed so far: Schwarzschild (left), Gullstrand--Painlevé (middle), and Eddington--Finkelstein (right). The upper panel shows the propagation of null rays on the coordinate grid, while the lower panel displays the corresponding characteristic speeds. In Schwarzschild time, both the ingoing and outgoing speeds vanish at the Schwarzschild radius. In GP and EF time, only the outgoing speed (represented by the dashed line) vanishes at the event horizon, illustrating its nature as a ``one-way membrane" \cite{finkelstein_past-future_1958}.

\subsection{Maximal analytic extension}

Despite Lemaître's recognition of the fictitious nature of the Schwarzschild singularity, it was not until the late 1950s that relativists reached a consensus on its nature. An essential role was played by the Kruskal--Szekeres (KS) coordinates, based on a double-null form of the Schwarzschild metric, providing a unique, maximal, analytic extension of Schwarzschild spacetime. These coordinates were published independently by Kruskal \cite{kruskal_maximal_1960} and Szekeres \cite{szekeres1960singularities, szekeres_golden_2002} in 1960, with earlier contributions by Synge in 1950 \cite{synge_gravitational_1950} and Fronsdal in 1959 \cite{fronsdal_completion_1959}.

\begin{figure}[h]%
    \centering
    \includegraphics[height=0.33\textheight]{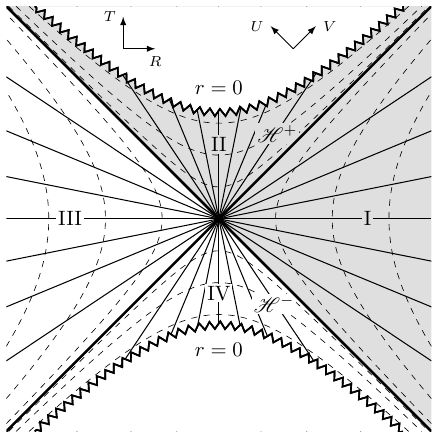}
    \caption{Maximally extended Schwarzschild geometry in Kruskal--Szekeres coordinates. Constant lines of Schwarzschild coordinates are depicted as solid, straight lines for $t$ and dashed, hyperboloidal lines for $r$. Extension in EF and GP coordinates covers the domain shaded in gray.}
    \label{fig:maximal}
\end{figure}

The original paper by Kruskal\footnote{Israel reports in \cite{israel1987dark} that Kruskal---a mathematician and plasma physicist---derived the maximal extension of Schwarzschild spacetime by following the path that leads from Rindler to Minkowski coordinates while learning general relativity in a small study group at Princeton in the mid-1950s. The paper \cite{kruskal_maximal_1960} was apparently written by Wheeler with the author's credit given to Kruskal.} is a short description of the coordinates,
\be\label{eqn:kruskal} 
T = e^{\frac{r}{4M}} \sqrt{\frac{r}{2M} - 1}\ \cosh \frac{t}{4M},\qquad  R = e^{\frac{r}{4M}} \sqrt{\frac{r}{2M} - 1} \ \sinh \frac{t}{4M}. 
\ee
The coordinates, defined for $r>2M$, can be extended to the full range $r>0$. The resulting Kruskal diagram of the maximally extended Schwarzschild spacetime is illustrated in Fig.~\ref{fig:maximal}. It is instructive to list the steps in some detail because the interplay of regularization and extension based on null rays is relevant for hyperboloidal coordinates. The transformations only act on coordinates $t$ and $r$. Therefore, we restrict our discussion to the $tr$-plane.
\begin{itemize}
    \item[(i)] Write the Schwarzschild metric \eqref{eqn:schwarzschild} in the tortoise coordinate \eqref{eqn:tortoise},
        \begin{equation}
        r_\ast = \int \frac{dr}{f},  \qquad ds^2 = f \left(-dt^2 + dr_\ast^2\right).
        \label{eqn:tortoise_schwarzschild}
        \end{equation}
    \item[(ii)] Transform to double-null coordinates,
        \begin{equation}
        u = t-r_\ast, \quad v = t+r_\ast, \qquad ds^2 = -f \, du \, dv.
        \label{eqn:double_null}
        \end{equation}
    \item[(iii)] Transform to extensible coordinates, resulting in a regular metric across the horizon,
        \begin{equation}
        U = -e^{-\frac{u}{4M}}, \quad V = e^{\frac{v}{4M}},  \qquad ds^2 = -\frac{32 M^3}{r} \exp\left(-\frac{r}{2M} \right) dU \, dV.
        \label{eqn:extensible_coords}
        \end{equation}
    \item[(iv)]  Transform back to time and space coordinates,
        \begin{equation}
        T = \frac{1}{2} (V+U), \quad R = \frac{1}{2} (V-U), \qquad ds^2 = \frac{32 M^3}{r} \exp\left(-\frac{r}{2M} \right) \left(- dT^2 + dR^2 \right).
        \label{eqn:time_space_coords}
        \end{equation}
\end{itemize}


Figure \ref{fig:maximal} presents the maximally extended Schwarzschild spacetime in Kruskal--Szekeres coordinates. The expressions \eqref{eqn:kruskal} are only defined in region I for $T > 0$ and $R > 0$, but can be extended to other regions bounded by the past and future singularities at $T^2 - R^2 = 1$, corresponding to the curvature singularity at $r = 0$. The figure also shows level sets of Schwarzschild--Droste coordinates: solid straight lines for $t$ and dashed hyperboloidal lines for $r$. The Schwarzschild time slices intersect at the bifurcation sphere, visually demonstrating the coordinate singularity. In contrast, KS time slices are horizon-penetrating. 

The hyperbolic nature of the coordinate transformation near the horizon is clear from the use of hyperbolic functions in the combined transformation \eqref{eqn:kruskal}.
The KS transformation regularizes the event horizon as follows. The first step (i) brings the Schwarzschild metric on the $tr$ plane to a conformally flat form. From this, we can immediately read off the incoming and outgoing null coordinates used in step (ii). We can write the tortoise coordinate as,
\begin{equation} \label{eqn:tortoise_k}
    r_\ast = r + \frac{1}{2\kappa_b} \left(\ln \frac{r}{r_b} + \ln |f|\right).
\end{equation}
The difference between the tortoise coordinate $r_\ast$ and the Schwarzschild areal coordinate $r$ is given by two logarithmic terms that blow up at either end: one toward infinity, and another toward the horizon. The null coordinates defined in step (iii) are crucial for the maximal extension beyond the horizons, and can be written as $U = -e^{-\kappa_b u}$ and $V = e^{\kappa_b v}$. The final step (iv), which introduces space and time coordinates, may seem cosmetic but reveals the hyperbolic nature of the transformations. Finally, note that one can construct maximal extensions using a double-time coordinate instead of a double-null coordinate by considering ingoing and outgoing versions of horizon-penetrating coordinates \cite{lemos_maximal_2021}.

\subsection{Conformal compactification}
\label{sec:2_conformal}
The Schwarzschild event horizon is a limiting surface of no return and it is not the only such horizon in Schwarzschild spacetime. Horizons mark the boundary of accessible regions, defining what can and cannot be observed or interacted with. When we assume that no information comes in from infinity, we treat the asymptotic region as another horizon or boundary.

The global causal structure of spacetimes was only understood after Penrose introduced conformal compactification to general relativity in 1963 \cite{penrose_asymptotic_1963}. This groundbreaking idea maps spacetime to a bounded domain, allowing a detailed study of the asymptotic behavior of propagating fields using local analysis. Conformal compactification also enabled illustrations of the global structure of spacetimes via simple diagrams. Carter \cite{carter1966complete} refined these conformal diagrams into the form we employ today.\footnote{These diagrams are known under many names such as conformal, Penrose, Penrose-Carter, or Carter-Penrose diagrams.}

\begin{figure}[h]
    \centering
    \includegraphics[height=0.3\textheight]{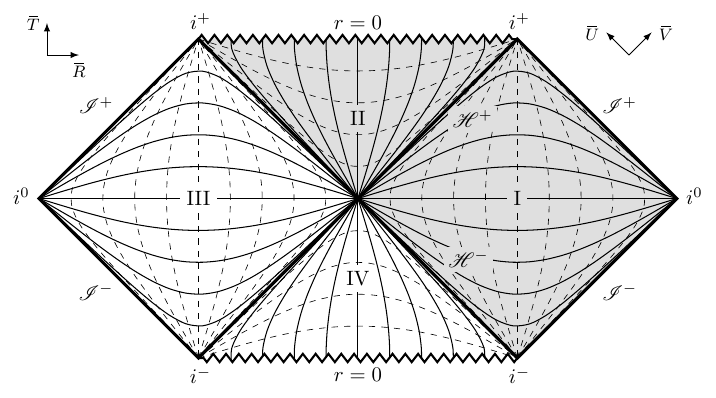}
    \caption{Maximally extended Schwarzschild geometry in Penrose coordinates. Constant lines of Schwarzschild--Droste coordinates $t$ and $r$ are depicted as solid and dashed lines, respectively. The shaded region is the same as in Fig.~\ref{fig:maximal}. Note the intersection of Schwarzschild time slices at the bifurcation sphere \textit{and} spatial infinity.}\label{fig:penrose}
\end{figure}

Conformal diagrams have two essential properties. First, in- and outgoing null rays are represented as straight lines at 45 degrees to the horizontal. Second, the full range of null rays is represented in a bounded domain. Compactification along null directions allows for representing the entire spacetime in a finite region. We start at the extensible double-null coordinates \eqref{eqn:extensible_coords} in step (iii) of the Kruskal--Szekeres construction, and map their infinite range to a finite one by using the tangent function: $\bar{U} = \arctan U$, $\bar{V} = \arctan V$. Other mappings are also possible. The barred coordinates have a bounded range, $\bar{U}, \bar{V} \in (-\pi/2, \pi/2)$. We then introduce space and time coordinates via $\bar{T} = \bar{V} + \bar{U}$, $\bar{R} = \bar{V} - \bar{U}$. The resulting conformal diagram for the maximally extended Schwarzschild spacetime is given in Fig.~\ref{fig:penrose}.

The construction of Penrose diagrams leads to a time coordinate that is horizon-penetrating near the black hole and hyperboloidal near infinity. It is these properties that make Penrose diagrams so useful in depicting the global causal structure. The compactification is not performed along Kruskal--Szekeres coordinates $\{T, R\}$ because they lack the appropriate asymptotic behavior near infinity. The Kruskal--Szekeres time $T$ is only horizon-penetrating, not hyperboloidal. Level sets of Penrose time $\widebar{T}$, on the other hand, are both horizon-penetrating and hyperboloidal, except for the values $0$, $\pi/2$, and $-\pi/2$. The level set $T = 0$ is a Cauchy surface, and the level sets $T = \pm\pi/2$ represent the future and past timelike infinities. The Penrose time in the range $|T| \in (0, \pi/2)$ is the first example of a horizon-penetrating, hyperboloidal time constructed in a black hole spacetime. Because of the essential role such hypersurfaces play in black-hole physics, both in asymptotically flat and asymptotically de Sitter spacetimes, we refer to them as \textit{bridges} (Sec.~\ref{sec:5}).

\subsection{Lessons from history}
\label{sec:lessons}

For nearly half a century, the true nature of the Schwarzschild radius was the subject of intense debate. The inadequacy of Schwarzschild time for describing across-horizon dynamics is evident in the fact that a particle falling into a black hole takes infinite Schwarzschild time, yet only a finite proper time. The coordinate velocity of a freely falling particle or an ingoing null ray approaches zero towards the horizon, rendering Schwarzschild time unsuitable for describing dynamics across the event horizon and conflicting with our modern understanding of a black hole. Today, we recognize the  ``magic'' sphere at $r_b=2M$ \cite{eddington1921space} as a mere coordinate singularity. Horizon-penetrating coordinates, such as GP \eqref{eqn:gp_trafo} or EF time \eqref{eqn:ef_trafo}, show that the event horizon is not a barrier for one-way propagation of matter and radiation. The velocity of an infalling particle is smooth across the horizon—a property not properly represented in Schwarzschild time. While general covariance guarantees equivalence among charts connected by diffeomorphisms, such equivalence is not valid for singular transformations. We must adapt the definition of time to the physically expected behavior.

The same principle applies to future null infinity, $\scri^+$. Outgoing null rays take infinite Schwarzschild time to reach $\scri^+$. This property was originally not considered problematic because $\scri^+$ is infinitely far away in proper distance. However, while massive particles do not reach $\scri^+$, null rays do. The conformal diagram in Fig.~\ref{fig:penrose} shows that the region near the bifurcation sphere is similar to the region around spatial infinity: both resemble Rindler wedges. Just as we use horizon-penetrating coordinates to describe across-horizon dynamics, we need coordinates that are regular across $\scri^+$ to describe the measurement of gravitational radiation by idealized observers.

Penrose coordinates satisfy these desired geometric properties: they are horizon-penetrating and hyperboloidal, bridging the black-hole event horizon with null infinity. They were not widely adopted in quantitative calculations because they break the time-translation symmetry of Schwarzschild spacetime. We need coordinates that are similar to GP and EF: break time-\emph{reflection} symmetry to ensure regularity across null horizons, but preserve time-\emph{translation} symmetry of the stationary background. Achieving this not only near the black-hole event horizon but also near null infinity is the topic of the next section.




\section{Hyperboloidal time near null infinity}\label{sec:3}

In the previous section, we saw that the regularization of the time coordinate near the black hole horizon resolves the intersection of time slices at the bifurcation sphere. Penrose's work revealed that standard time slices intersect also in the asymptotic region at spatial infinity (see Fig.~\ref{fig:penrose}). This coordinate singularity is independent of the presence of a black hole. All asymptotically flat spacetimes exhibit this behavior. In this section, we discuss the regularization of the time coordinate near infinity for flat spacetime, which provides the asymptotic structure for isolated systems.

As a first step, consider the Penrose coordinates $\{T,R\}$ in Minkowski spacetime. We compactify null coordinates $u=t-r$ and $v=t+r$ by applying a mapping such as the inverse tangent function, $U=\arctan u$ and $V=\arctan v$. The resulting Minkowski metric is singular at the boundaries,
\begin{align*}
    ds^2 &= -dt^2 + dr^2 + r^2 d\omega^2 = -dudv + \frac{1}{4}(v-u)^2d\omega^2 = \\
    &=\frac{1}{4 \cos^2 U\, \cos^2 V} \left( - 4 dUdV + \sin^2(V-U) d\omega^2 \right)
\end{align*}
The singularity can be captured by a conformal rescaling, $d\bar{s}^2 = \Omega^2 ds^2$ with $\Omega=2 \cos U \cos V$, so that the conformal metric $d\bar{s}^2$ is regular at infinity, where $\Omega=0$. Intuitively, one can think of this conformal regularization as taking care of the increase in metric distances by a decrease in the scaling such that the limit is regular. We then introduce time and space coordinates, $T=V+U$ and $R=V-U$, to obtain,
\be
\label{eqn:penrose_metric}
d\bar{s}^2 = - 4 dUdV + \sin^2(V-U) d\omega^2 = -dT^2 + dR^2 + \sin^2 R\, d\omega^2.
\ee

\begin{figure}[htb]
    \centering
    \subfloat[Minkowski]{\includegraphics[width=0.21\textwidth]{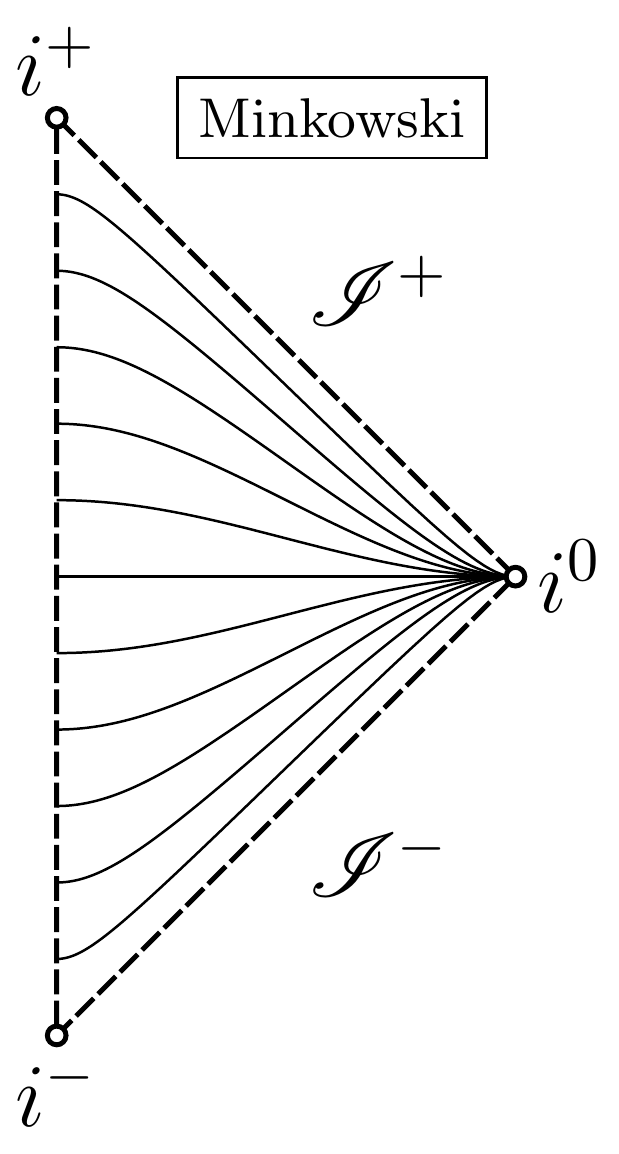}\label{fig_sub:minkowski}} \hfill
    \subfloat[Milne]{\includegraphics[width=0.21\textwidth]{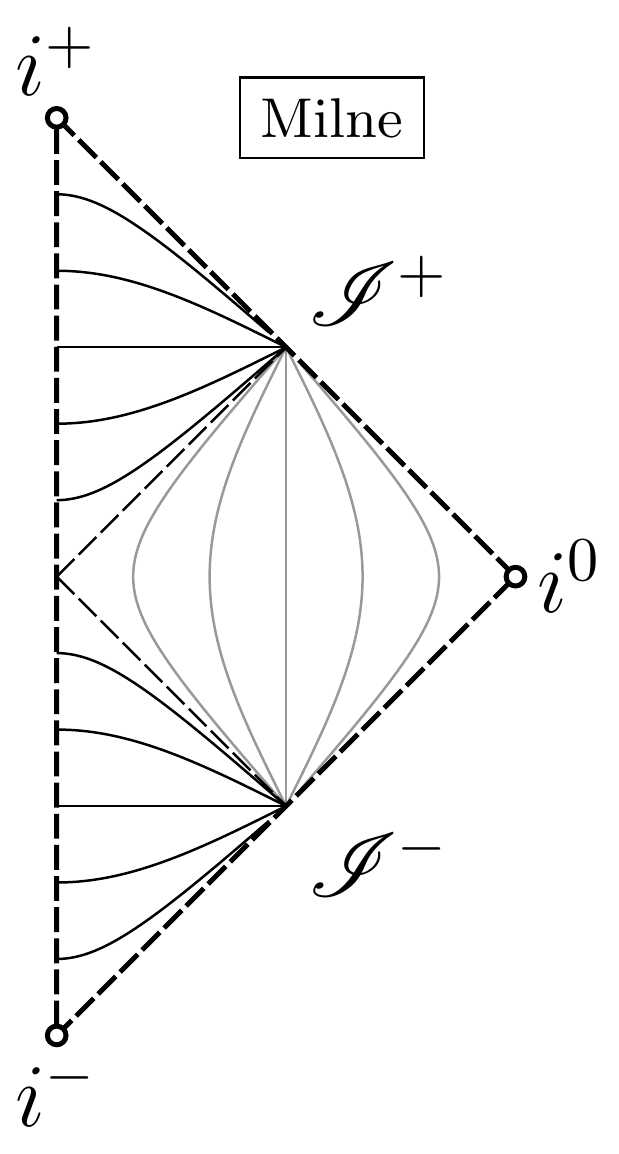}\label{fig_sub:milne}} \hfill
    \subfloat[Penrose]{\includegraphics[width=0.21\textwidth]{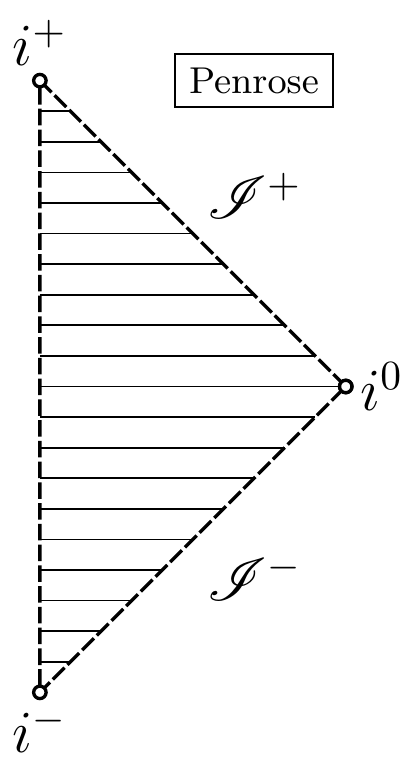}\label{fig_sub:penrose}} \hfill
    \subfloat[Hyperboloid]{\includegraphics[width=0.21\textwidth]{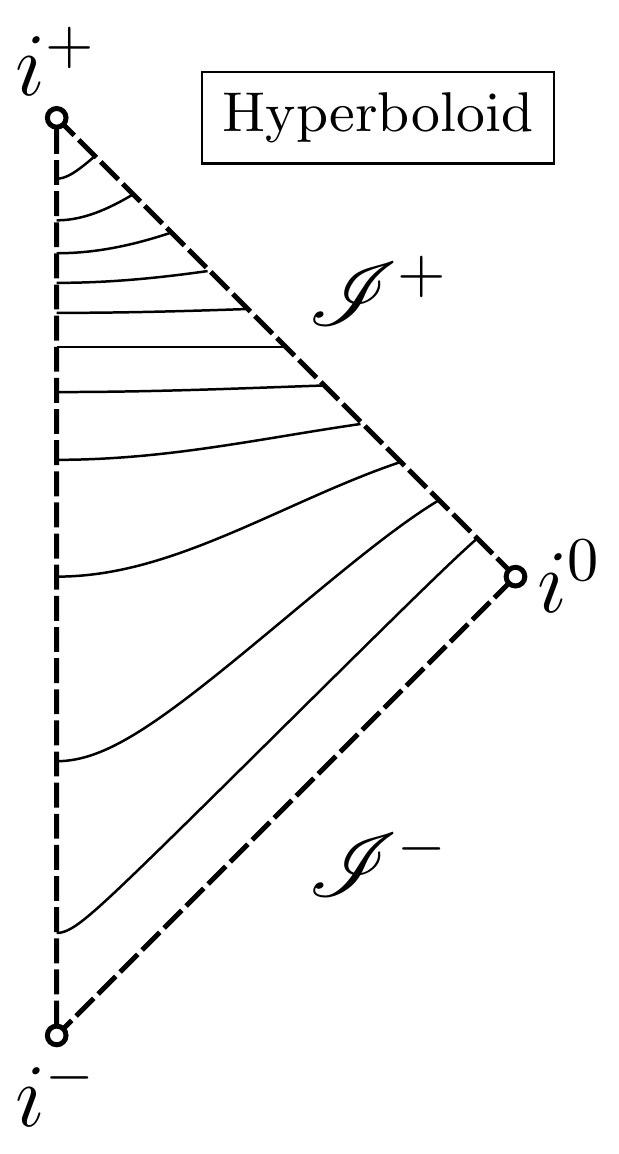}\label{fig_sub:gowdy}}
    \caption{The conformal diagrams of Minkowski spacetime depict the standard time coordinate (a), and three different hyperboloidal slicings: Milne (b), Penrose (c), and the hyperboloid slicing (d). The standard time slices intersect at spatial infinity. Milne's hyperbolic slices intersect at null infinity. In contrast, Penrose time slices and time-translated hyperboloids provide a smooth foliation of null infinity.}
    \label{fig:mink_penrose}
\end{figure}

The resulting conformal diagram of Minkowski space in Fig.~\ref{fig_sub:minkowski} shows that the level sets of Minkowski time $t$ intersect at spatial infinity. Historically, this intersection of time slices was not considered problematic because infinity is infinitely far away, in contrast with a black hole, where a free-falling observer can reach the event horizon in finite proper time. Observers cannot reach infinity by free-fall\footnote{While this property highlights the physical difference between horizons and null infinity, it also reveals a global geometric analogy: timelike geodesics near an event horizon have a similar causal structure as constantly accelerated timelike curves near null infinity, which underlies phenomena such as the Unruh effect and Hawking radiation \cite{jacobson2005introduction}.}. Therefore, the singularity at spatial infinity is somewhat out of sight.

However, radiation does reach null infinity at finite ``time". Time stops along null rays so the theory allows for light to travel to null infinity in a delicate limit revealed by conformal compactification. Idealized observers marching on null infinity along the integral curves of the asymptotic timelike Killing field can measure this radiation. The time surfaces they construct are necessarily hyperboloidal. Therefore, to study propagating waves in unbounded domains, it is helpful to construct a regular description of time at infinity. In principle, this problem extends beyond gravitational waves or black-hole perturbation theory to all wave phenomena, including acoustic or electromagnetic scattering in unbounded domains.

\subsection{Historical context for hyperboloidal time}

The importance of hyperbolic geometry in spacetime was already evident to Minkowski \cite{gray_non-euclidean_1999, galison2010minkowski}. In a famous talk in 1908, where Minkowski revealed the unification of space and time into spacetime, he emphasized the key role that the spacetime hyperboloid plays as the invariant surface under Lorentz transformations \cite{minkowski_1908}. Minkowski died within months of this talk and the hyperbolic approach became a niche topic. Over the following hundred years, time hypersurfaces with asymptotically hyperbolic geometry frequently appeared in the literature. Below I review the historical developments to explain some of the subtleties involved in choosing hyperboloidal time functions.

\subsubsection{Milne time}
\label{sec:milne}

In 1933, Milne proposed a cosmological model that turned out to be Minkowski spacetime in different coordinates \cite{milne1933world, possel2019teaching}. In Milne's model, spacetime hyperbolas (pseudo-hyper-spheres) serve as hypersurfaces of distant simultaneity and their radii serve as a time function. Milne coordinates resemble spherical coordinates based on spheres of different radii. Many early works exploiting the hyperbolic nature of the resulting time slices use this approach \cite{dirac1949forms, chen_hyperboloidal_1971, strichartz_harmonic_1973, fubini1973new, disessa1974quantization, sommerfield_quantization_1974, hostler1978coulomb}. 

Milne coordinates rely on a spacetime hyperboloid of radius $\eta$ given as the set of equidistant points from the origin,
\be\label{eqn:milne_slices}
t^2 - r^2 = \eta^2.
\ee
In Milne's hyperbolic slicing, the radius of these hyperboloids, $\eta$, acts as the time coordinate. Defining the comoving coordinate $\chi$ via $t=\eta \cosh\chi$ and $x = \eta \sinh\chi$, the Minkowski metric becomes
\be\label{eqn:milne} ds^2 = -d\eta^2 + \eta^2 \left(d\chi^2 +  \sinh^2 \chi d\omega^2 \right) . \ee
The level sets of $\eta$ are spaces of constant negative curvature. Their geometry matches exactly the geometry of standard time slices in Anti-de Sitter spacetime. It is this property that makes Milne slicing appealing in recent works on quantum field theory and flat-space holography \cite{de_boer_holographic_2003, campiglia2015asymptotic, cheung_4d_2017, strominger2018lectures, raclariu2021lectures, ogawa2023wedge, chen2024entanglement, sleight2024celestial}.

The main problem with Milne slicing is its time dependence. Milne time breaks the time-translation symmetry of Minkowski spacetime, leading to a time-dependent metric \eqref{eqn:milne}. In addition, the slices intersect at null infinity and do not provide a smooth foliation of the conformal boundary (see Fig.~\ref{fig_sub:milne}). This problem is similar in nature to the intersection of Schwarzschild slices at the bifurcation sphere and the intersection of standard Minkowski time slices at spatial infinity. Time freezes at infinity and we get a singular coordinate at the conformal boundary. 

Milne slicing demonstrates that approaching null infinity does not necessarily provide a good description for spacetime or the asymptotic region. We must also ensure that the slicing provides a faithful definition of time at the asymptotic boundary. What we need is a time coordinate that (i) provides a non-intersecting foliation of the conformal boundary, and (ii) respects the time-translation symmetry of the underlying Minkowski spacetime. The first condition was satisfied by Penrose time.



\subsubsection{Penrose time}
\label{sec:penrose}

Level sets of Penrose time $T$ used in metric \eqref{eqn:penrose_metric} are depicted in  Fig.~\ref{fig_sub:penrose}. Penrose time provides a non-intersecting foliation of null infinity, satisfying our first requirement for a good time coordinate in Minkowski spacetime. The connection of Penrose time to hyperboloidal coordinates is not immediately obvious. The standard derivation of Penrose time involves compactification along null directions. When written in terms of the Minkowski coordinates, however, we see that Penrose time is hyperboloidal \cite{zenginouglu2024hyperbolic}. We can write Penrose time as
\[ T = V+U = \arctan v + \arctan u = \arctan \frac{v+u}{1-uv} = \arctan \frac{2 t}{1-\left(t^2-r^2\right)}. \]
Defining $\tilde{T}\equiv -1/\tan T$, we get
\be\label{eqn:penrose_time} t = \tilde{T} \pm \sqrt{r^2 + \tilde{T} + 1}. \ee
Compare this expression to Milne time with $t = \pm \sqrt{r^2 + \eta^2}$. The crucial difference is the shift by $\tilde{T}$ on the right hand side. The shift allows Penrose time to foliate null infinity smoothly. The time-dependence inside the square root allows the coordinates to cover past null infinity, the Cauchy surface at $t=0$, and future null infinity. This global coverage makes Penrose coordinates ideal to depict the causal structure, but their time-dependence makes them cumbersome for practical computations.
If one is interested primarily in measurements along future null infinity, introducing such time dependence into the description unnecessarily complicates the calculations performed in a stationary background. This problem was solved by Gowdy \cite{gowdy_wave_1981} based on the hyperboloid slicing \cite{smarr_kinematical_1978}.

\subsubsection{Hyperboloid slicing}
\label{sec:hyperboloid}


In 1978, Smarr and York studied various time slicing conditions for solving Einstein equations \cite{smarr_kinematical_1978}. Among the ``promising" slicings they analyzed was the hyperboloid slicing with constant mean extrinsic curvature. One way to construct these slices is to fix a hyperboloid with a constant radius of curvature (set $\eta=L$ in \eqref{eqn:milne_slices}) and shift it along the time direction by $\tau$:
\be
\label{eqn:gowdy_tau} (t-\tau) - r^2 = L^2, \quad \mathrm{or} \quad t = \tau \pm \sqrt{r^2+L^2}.
\ee
Here, $L$ is the radius of curvature of the hyperboloid with constant mean extrinsic curvature, $K=3/L$. Compare this expression to Penrose time in \eqref{eqn:penrose_time}. The spacetime hyperboloids are shifted along the timelike Killing field without a time-dependence in their radius. Using the coordinate $\chi$ with $r=L \sinh\chi$, we can write the Minkowski metric as
\begin{align*} ds^2 &= -d\tau^2 \mp \frac{2r}{\sqrt{L^2+r^2}} d\tau dr + \frac{L^2}{L^2+r^2} dr^2 + r^2 d\omega^2 \\ 
& = -d\tau^2 \mp 2 L \sinh \chi d\tau d\chi + L^2 \left(d\chi^2 + \sinh \chi d\omega^2\right) . 
\end{align*}
Comparing this expression with its Milne counterpart \eqref{eqn:milne}, we see that the spatial hypersurfaces have the same geometry: hyperbolic spaces with constant curvature. The hyperboloid slicing, however, is more natural for a description of spacetime from the point of view of an idealized asymptotic observer. In hyperboloid slicing, time at null infinity is not frozen but flows. The slices do not intersect at infinity and do not single out a point in spacetime. In situations where a hyperbolic geometry for spatial hypersurfaces may be preferable, hyperboloid slicing is more natural than Milne slicing.

However, the Minkowski metric in hyperboloid slicing is not suitable for numerical calculations because lapse and shift are unbounded towards infinity. In the ADM decomposition of a spherically symmetric metric, we write
\[ ds^2 = \left(-\alpha^2 + \gamma^2 \beta^2 \right)dt^2 + 2 \gamma^2 \beta dt dr + \gamma^2 dr^2 + r^2 d\omega^2. \]
We read off lapse $\alpha$, shift $\beta$, and the metric function $\gamma$ as 
\[ \alpha = \sqrt{1+\frac{r^2}{L^2}}, \qquad \beta = -\frac{r}{L} \alpha, \qquad \gamma=\frac{1}{\alpha}. \]
The lapse grows as $\alpha\sim \mathcal{O}(r)$ and the shift grows as $\beta \sim \mathcal{O}(r^2)$. 
As a consequence, the time step in free evolutions become prohibitive toward infinity and null infinity cannot be included in the computational domain (for an attempt to use such a foliation in a black hole spacetime, see \cite{gentle_constant_2001}).

The solution to this problem is compactification. Such compactified coordinates were introduced by Gowdy in 1981 \cite{gowdy_wave_1981}. Gowdy's motivation was to follow gravitational waves ``all the way to future null infinity in a finite number of steps on a finite grid." For this purpose, compactification along the radial direction is indispensable. Gowdy introduces a ``wave distance function" preserving the coordinate expression of outgoing characteristics. Using $\sigma$ for the compactifying radial coordinate and $\tau$ for the hyperboloidal time coordinate, Gowdy's wave distance function satisfies $t-r = \tau + \sigma$. This expression is similar to the EF coordinates satisfying \eqref{eqn:ef_trafo}. Aside from a change in sign due to flipping the locations of the origin and infinity, the construction is characteristic-preserving in our terminology. The compactifying coordinate $\sigma\in[0,1]$ is given by
\be
\label{eqn:gowdy_sigma}
\sigma = -\frac{r}{L} + \sqrt{1+\frac{r^2}{L^2}}, \quad \mathrm{or} \quad r = \frac{L}{2}\left(\frac{1}{\sigma} - \sigma\right).
\ee
Choosing the conformal factor $\Omega=\sigma/L$, the conformally rescaled Minkowski metric becomes
\[ d\bar{s}^2 = - \frac{\sigma^2}{L^2} d\tau^2 + \frac{1-\sigma^2}{L} d\tau d\sigma + d\sigma^2 + \frac{(1-\sigma^2)^2}{4} d\omega^2. \]
Gowdy arrived at these coordinates using a conformal projection but one can generalize this procedure for different choices of compactifications (see Sec.~\ref{sec:spatial_compactifications}). About twenty years after Gowdy's work, Moncrief used hyperboloid slicing in a talk on a conformal approach to Einstein equations for numerical relativity \cite{Moncrief00} using a different radial compactification function,
\be\label{eqn:Moncrief_rho} r = \frac{2 L \rho}{1-\rho^2}, \qquad \mathrm{or} \qquad \rho = - \frac{L}{r} + \sqrt{1+\frac{L^2}{r^2}}. \ee
Choosing the conformal factor $\Omega=(1-\rho^2)/2L$, the conformally rescaled Minkowski metric becomes
\be\label{eqn:mink_spatially_flat} d\bar{s}^2 = \frac{(1-\rho^2)^2}{4L^2} d\tau^2 - \frac{2\rho}{L} d\tau d\rho + d\rho^2 + \rho^2 d\omega^2. \ee
The hypersurfaces of the conformal metric are spatially flat. We see that, using the same hyperboloid slicing, we can choose a conformal compactification that is characteristic-preserving (as given by Gowdy in \cite{gowdy_wave_1981}) or spatially flat (as given by Moncrief in \cite{Moncrief00}). Either way, a compactification using hyperboloid slicing satisfies our requirements for a suitable time function for numerical calculations. While Gowdy's paper was ignored, the spatially flat representation by Moncrief led to the first numerical studies using hyperboloidal foliations in Minkowski spacetime \cite{husa2003numerical, FodorRacz04, fodor2008numerical, bizon2009universality}.

The hyperboloid slicing and the conformal compactification provides a prototype for a succesful implementation of the hyperboloidal method. Next, we generalize the hyperboloid slicing to explore different aspects of hyperboloidal compactification in Minkowski spacetime. 

\subsection{Generalization}
In the previous section, we resolved the intersection of standard time slices at spatial infinity by switching to a time coordinate that foliates null infinity. The level sets of such time coordinates have asymptotically hyperbolic geometry. However, this property by itself is not sufficient for a good time function as demonstrated by Milne time (\ref{sec:milne}). What we need is a non-intersecting coordinate similar to Penrose time (\ref{sec:penrose}) but is stationary, which led us to the hyperboloid slicing with constant mean curvature hypersurfaces (\ref{sec:hyperboloid}). Compactification along such hypersurfaces gives us a suitable representation of Minkowski spacetime for computations of outgoing radiation.

In summary, there are two essential ingredients to hyperboloidal compactification: (i) the choice of a hyperboloidal time function, and (ii) the choice of a compactifying coordinate. For the hyperboloid slicing, we used the time-shifted hyperboloids \eqref{eqn:gowdy_tau} with the characteristic-preserving compactification by Gowdy \eqref{eqn:gowdy_sigma} or the spatially flat one by Moncrief \eqref{eqn:Moncrief_rho}. In this section, we generalize this procedure while maintaining its essential properties for numerical and analytical studies of waves. We start with a few useful definitions.

\subsubsection{Definitions}\label{sec:definitions}

The Penrose compactification procedure suggests a geometric definition for asymptotically flat or asymptotically simple spacetimes \cite{frauendiener2004conformal}. For our discussion, it is helpful to start with the Riemannian geometry of spatial hypersurfaces representing moments of time. We consider a smooth Riemannian manifold $M$ of dimension \( n \) without boundary, equipped with a Riemannian metric \(g\). We carve spacetime into slices represented by such hypersurfaces with suitable properties, such as conformal compactness and asymptotic hyperbolicity \cite{chrusciel2003mass, Cederbaum_2015, Beyer_2022}.


\begin{definition}
    A smooth Riemannian manifold of dimension $n$ without boundary \((M, g)\) is  \textbf{conformally compact} if the following conditions are satisfied:
    \begin{enumerate}
        \item There exists a compact manifold-with-boundary \(\overline{M}\) such that \(M\) is diffeomorphic to the interior of \(\overline{M}\).
        \item There exists a smooth conformal factor (also called defining function) \(\Omega\) such that:
        \begin{itemize}
            \item \(\Omega > 0\) on \(M\),
            \item \(\Omega = 0\) on \(\partial \overline{M}\), and
            \item \(\mathrm{d}\Omega \neq 0\) on \(\partial \overline{M}\).
        \end{itemize}
        \item The \textit{conformally rescaled metric} \(\widetilde{g} = \Omega^2 g\) extends smoothly to all of \(\overline{M}\).
    \end{enumerate}
\end{definition}

Intuitively, we think of the conformal boundary \( \partial\overline{M} \) as representing the points at infinity. The conformal factor $\Omega$ goes to zero at the boundary reflecting that the original metric blows up near infinity. Note that the standard Euclidean space is not conformally compact whereas hyperbolic space is. 

The three-dimensional metric describing the geometry on the level sets of Penrose time $T$ is conformally compact \eqref{eqn:penrose_metric}. In addition to conformal compactness, we would like to impose further restrictions on the behavior of the curvature near the asymptotic boundary. 

\vspace{3mm}

\begin{definition}
    Let \( (M, g) \) be conformally compact with conformal boundary \( \partial \overline{M} \) and conformal factor \( \Omega \). The manifold \( (M, g) \) is said to be \textbf{asymptotically hyperbolic} if the Ricci scalar of the metric approaches the Ricci scalar of hyperbolic space:
        \[ R_g = - n(n-1) + O(\Omega^2) \quad \mathrm{as} \quad \Omega \to 0. \]
    In particular, the metric asymptotically approaches the hyperbolic metric in Fefferman--Graham form \cite{fefferman1985conformal, fefferman2012ambient} with
    \be\label{eqn:fg_form} g = \frac{L^2}{r^2} dr^2 + r^2 d\omega^2, \quad \mathrm{as} \quad r\to\infty.\ee
\end{definition}

These definitions serve to clarify the nature of the induced metric toward the conformal boundary. Consider the Minkowski metric in Penrose coordinates given in \eqref{eqn:penrose_metric}. In the compactified double-null coordinates $(U,V)$, the induced metric on level sets of $U$ is degenerate. In contrast, the Penrose coordinates $(T,R)$ lead to an induced metric that is  asymptotically hyperbolic. The conformal metric in Penrose coordinates is the metric of the Einstein cylinder. Many other choices are possible. For example, we have seen that combining the hyperboloid slicing of Sec.~\ref{sec:hyperboloid} with a suitable choice of conformal compactification leads to a spatially flat induced metric (see \eqref{eqn:mink_spatially_flat}). Among these qualitatively similar choices, the essential geometric requirement is that the induced metric is asymptotically hyperbolic. When this condition is satisfied, the conformal boundary represents a cut of null infinity in the spacetime picture.

The definition of asymptotically hyperbolic manifolds allows us to distinguish characteristic foliations from hyperboloidal ones. However, the definition uses only the geometry of the spatial hypersurfaces. As the comparison with Milne coordinates from Sec.~\ref{sec:milne} shows, we also need to make sure that the conformal boundary is foliated by a stack of non-intersecting asymptotically hyperbolic manifolds. 

\begin{definition}\label{def:asymptotically_hyperbolic}
    Let \(\overline{M}\) be a \((3+1)\)-dimensional spacetime manifold with a conformal boundary \(\partial \overline{M}\). A one-parameter family of spacelike hypersurfaces \(\{\Sigma_\tau\} \subset \widehat{M}\) is \textbf{globally regular} if the associated lapse function \(\alpha\) is smooth and nowhere vanishing on \(\widehat{M}\), including at the conformal boundary \(\partial \widehat{M}\).
\end{definition}

This definition ensures that each hypersurface \(\Sigma_\tau\) is well-defined and non-intersecting throughout the entire extended spacetime, up to and including the boundary. In particular, it allows us to avoid slicings such as the Milne slicing that intersect at the conformal boundary. Note that both Milne coordinates and standard Minkowski coordinates break the condition of global regularity. Finally, we also want our foliation to respect the time-translation invariance of the background.

\begin{definition}
    Let \((M, g)\) be a spacetime admitting a timelike Killing vector field \(\xi^\mu\). 
    A foliation \(\{\Sigma_\tau\}\) of \(M\) is \textbf{stationary} if the associated time-evolution vector field \(\partial_\tau^\mu\) is everywhere proportional to \(\xi^\mu\), namely $\partial_\tau^\mu \;=\; f\,\xi^\mu$, with \(\mathcal{L}_\xi f = 0\) and \(f > 0\), ensuring \(\partial_\tau^\mu\) is timelike and future-directed.
\end{definition}

A stationary foliation if globally regular if any of its leaves approaches null infinity.

\subsection{Hyperboloidal time functions}\label{sec:general_time}
With the definitions from the previous section at hand, we can now specify the essential features of time-shifted hyperboloids \eqref{eqn:gowdy_tau}. The standard Minkowski foliation by level sets of $t$ is stationary but neither conformally compact nor globally regular. Therefore, Minkowski time is not suitable for studying asymptotic properties of radiative fields unless one is particularly interested in the structure of spatial infinity \cite{macedo2018spectral}. Milne time \eqref{eqn:milne_slices} has asymptotically hyperbolic slices but the foliation is neither stationary nor globally regular and time freezes at infinity. While the geometry of the spatial hypersurfaces has some useful properties, the slicing does not capture energy decay due to radiation. Penrose time \eqref{eqn:penrose_time} is globally regular and leads to asymptotically hyperbolic slices but is not stationary. From a causal perspective, the lack of stationarity is not a problem. In fact, it is this property that allows Penrose coordinates to cover the entire Minkowski spacetime, including the null infinities, the timelike infinities, and spatial infinity. A stationary foliation can only cover one of those regions due to the geometric properties of the timelike Killing field. However, a stationary foliation significantly simplifies the analyis of radiative fields. 

Among the examples we discussed, only the hyperboloid slicing in \eqref{eqn:gowdy_tau} is stationary and globally regular with asymptotically hyperbolic slices. The time-shifted hyperboloids represent an instant of time for the idealized observer at infinity. They allow us to describe the dynamical evolution of wave equations solving a hyperboloidal initial value problem \cite{friedrich2002conformal} and measure the energy carried out by outgoing radiation to future null infinity. The hyperboloid slicing is clearly not the only slicing of Minkowski spacetime satisfying these properties. A generalization of the slicing given in \eqref{eqn:gowdy_tau} can be achieved simply by choosing a time function $\tau$ via \cite{zenginoglu_hyperboloidal_2008},
\be\label{eqn:height_function} 
    \tau = t + h(r), 
\ee
where $h$ is referred to as the height function because it describes the ``height'' of the spacelike surface with respect to $t$ at each $r$ \cite{beig1998late}. The height function preserves the spherical and time-translation symmetries of the background because it is independent of the angular and time coordinates. Gowdy had already used the height-function approach in his work using the hyperboloid slicing \cite{gowdy_wave_1981}. The Minkowski metric reads
\begin{equation}
\label{eqn:theta_metric} 
ds^2 = -d\tau^2 + 2 H(r)\, d\tau dr + \left(1-H(r)^2\right) dr^2 + r^2 d\omega^2,
\end{equation}
where $H(r):=dh/dr$ is called the boost function. One can write down general conditions that the boost function must satisfy so that the time slices have asymptotically hyperbolic geometry \cite{zenginoglu_hyperboloidal_2008,zenginouglu2024hyperbolic}:
\begin{equation}\label{eqn:time_trafo} |H(r)|<1, \quad \forall r\in [0,\infty) \quad \& \quad |H(r)| = 1 - \frac{C}{r^2} + \mathcal{O}(r^{-3}), \  \mathrm{as}\ r\to\infty, \end{equation}
with some constant $C>0$. The first condition ensures the $\tau$-hypersurfaces are spacelike, providing a time function. The second condition ensures the $\tau$-hypersurfaces have asymptotically hyperbolic geometry. To understand the second condition better, consider that we get null coordinates for $H(r)=\pm 1$, or $h(r)=\pm r$. The nonvanishing $C$ is required to avoid degeneracy of the induced metric. If the boost function appraches $1$ more rapidly, we call the foliation asymptotically null. The constant $C$ determines how rapidly our foliation approaches null infinity. For small $C$, the foliation resembles a null foliation and has fast characteristics. For large $C$, the foliation resembles the standard time coordinate and has slow characteristics. The sign of $H$ determines whether the foliation is future-directed or past-directed, meaning whether $\tau$ asymptotes to outgoing null rays $u$ or incoming null rays $v$.

Hyperboloidal surfaces asymptote toward the null cone as $H\to 1$ so they are sometimes referred to as asymptotically null \cite{calabrese2006asymptotically, misner2006excising}. Gowdy refers to these coordinates as asymptotically retarded \cite{gowdy_wave_1981}. However, these hypersurfaces are spacelike everywhere, including asymptotically at null infinity. Following Friedrich \cite{Friedrich1983}, we refer to the generalization of hyperboloid slicing as hyperboloidal, defined as a globally regular, spacelike foliation that approaches null infinity. Note that the term hyperboloidal is used more broadly in the literature. Any surface that resembles a hyperboloid in certain coordinates can be called hyperboloidal. In particular, one could argue that horizon-penetrating surfaces are hyperboloidal in this broader use of the term. In this paper, we reserve the term hyperboloidal for foliations with asymptotically hyperbolic hypersurfaces as defined in Sec.~\ref{sec:definitions}. 

One reason hyperboloidal coordinates are inconvenient compared to the standard slicing is the off-diagonal term in \eqref{eqn:theta_metric}. This term is necessary to keep the time-translation symmetry of the background, just as with stationary, horizon-penetrating coordinates in \eqref{eqn:gp} or \eqref{eqn:ef}. As demonstrated by Lema\^itre \cite{lemaitre1933univers, lemaitre_expanding_1997}, the non-diagonal terms can be removed by employing the freedom in the spatial transformation. For the time transformation \eqref{eqn:height_function} with the differential $d\tau = dt + H dr$, we define a space coordinate $\xi$ via $d\xi = dt + \frac{1}{H} dr$.
The transformed metric is diagonal,
\[ ds^2 = \frac{1}{1-H^2} \left(-d\tau^2 + H^2 d\xi^2\right) + r^2 d\omega^2. \]
However, the metric is not stationary and therefore not as convenient for numerical computations.

\subsubsection{Examples}\label{sec:examples}

The class of stationary, globally regular time functions with asymptotically hyperbolic level sets is sufficiently large to adapt to various scenarios. This flexibility is useful for applications when used appropriately. Below we give a few examples to illustrate the properties of such stationary hyperboloidal coordinates. In these examples, consider hypersurfaces approaching future null infinity. The case for past null infinity follows with a change of sign for the height function. For practical applications, one must also consider parity requirements at the origin (see \cite{vano2024height} for more examples).

\begin{example}[Simplest example]
The simplest height function that satisfies the conditions \eqref{eqn:time_trafo} and is regular at the origin is
\be
\label{eqn:ha}
-h_a(r) = r + \frac{C}{1+r}, \qquad -H_a(r)= 1-\frac{C}{(1+r)^2},
\ee
\end{example}

\begin{example}[Source-adapted hyperboloid]
We can write the hyperboloid slicing \eqref{eqn:gowdy_tau} in a more general form by shifting its center:
\be
\label{eqn:hb}
-h_b(r) = \sqrt{L^2 + (r-r_t)^2}, \qquad -H_b(r)= \frac{r-r_t}{\sqrt{L^2 + (r-r_t)^2}}.
\ee
Such off-centered hyperboloids may help with the efficient computation of radiation emitted by a source near the turning point $r=r_t$. For example, in a black hole spacetime with a particle perturbation, $r_t$ would be the location of the particle that perturbs the background spacetime and $r$ would be the tortoise coordinate (see \eqref{eqn:sa_hypal}). This foliation smoothly transitions to the ingoing null cone for $r < r_t$ and to the outgoing null cone for $r > r_t$. Consequently, the dominant characteristic speeds of ingoing and outgoing rays exchange at $r_t$, which is the reason for the term turning point: the ingoing speed exceeds the outgoing speed for $r < r_t$, while the outgoing speed surpasses the ingoing speed for $r > r_t$. Since waves emanate from the source at $r_t$ in both directions, this foliation efficiently captures the propagation of these waves.
\end{example}

\begin{figure}
    \renewcommand\thesubfigure{\alph{subfigure}} 
    \centering
    \subfloat[]{\includegraphics[width=0.16\textwidth]{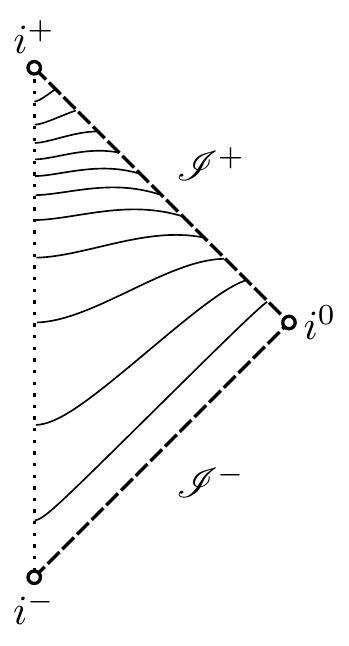}\label{fig:example_a}}
    \subfloat[]{\includegraphics[width=0.16\textwidth]{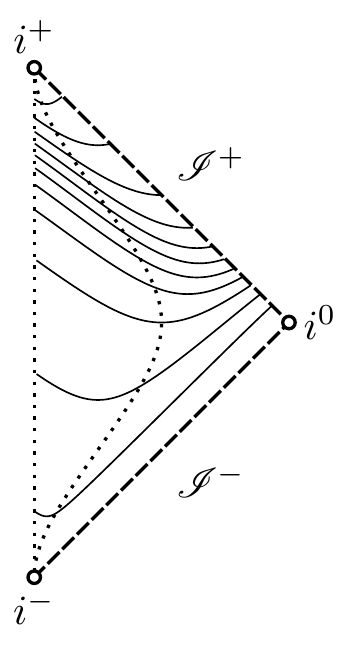}\label{fig:example_b}}
    \subfloat[]{\includegraphics[width=0.16\textwidth]{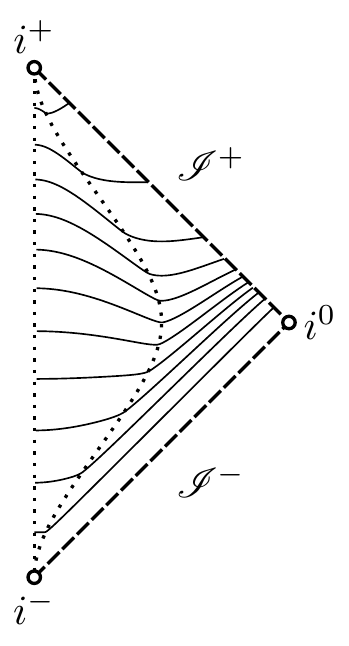}\label{fig:example_c}}
    \subfloat[]{\includegraphics[width=0.16\textwidth]{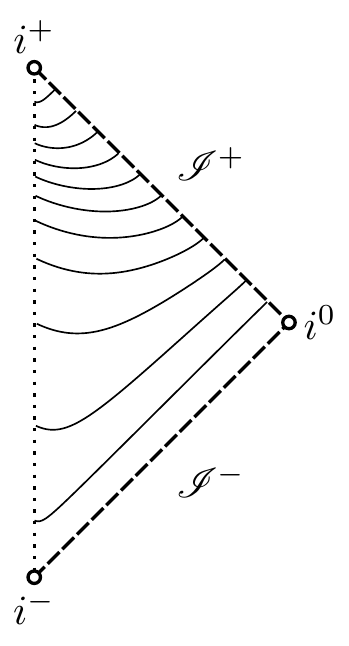}\label{fig:example_d}}
    \subfloat[]{\includegraphics[width=0.16\textwidth]{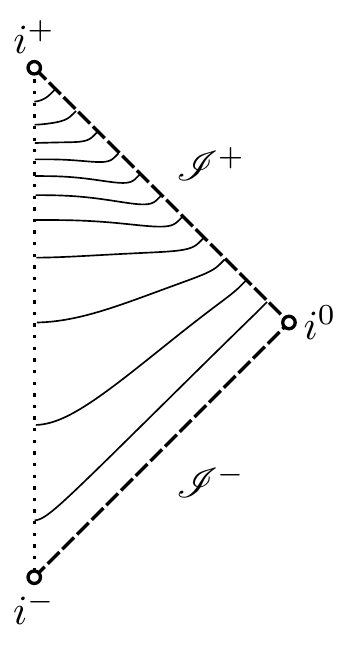}\label{fig:example_e}}
    \subfloat[]{\includegraphics[width=0.16\textwidth]{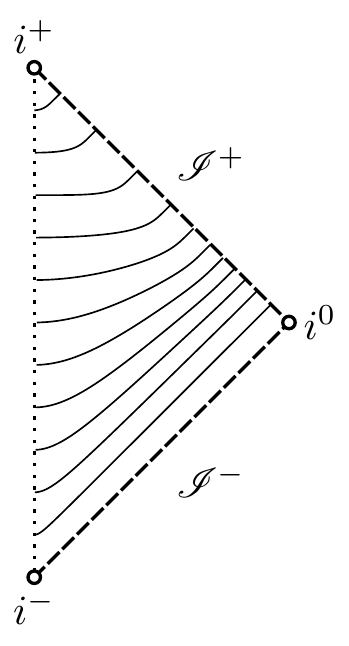}\label{fig:example_f}}
    \caption{Six examples of foliations corresponding to equations \eqref{eqn:ha} with $C=1$, \eqref{eqn:hb} with $L=1$ and $r_0=1$, \eqref{eqn:hc} with $L=1$, \eqref{eqn:hd}, \eqref{eqn:he}, and \eqref{eqn:hf}. The first three are hyperboloidal; the last three are asymptotically null in the Penrose compactification and therefore become diagonal at null infinity. In Figs.~\ref{fig:example_b} and \ref{fig:example_c}, the timelike surface $r=1$ is plotted as a dotted line.}
    \label{fig:examples}
\end{figure}

\begin{example}[Fefferman--Graham--Bondi]
In the definition of asymptotically hyperbolic Riemannian manifolds, we used the Fefferman--Graham form of a metric \eqref{eqn:fg_form}. The FG form of a spacetime metric can only be used for cosmological spacetimes with a non-vanishing cosmological constant \cite{compere2019lambda, ciambelli2024radiation}. However, for asymptotically flat spacetimes, we can bring the metric to a form that combines the properties of FG and Bondi gauges. We set,
\be\label{eqn:hc} 
-h_c(r) = \sqrt{r^2-L^2} - L \arctan\sqrt{\frac{r^2}{L^2}-1}, \qquad -H_c(r) = \sqrt{1-\frac{L^2}{r^2}}.
\ee
The Minkowski metric \eqref{eqn:theta_metric} becomes
\begin{equation}\label{eqn:fg_metric}
ds^2 = -d\tau^2 - 2 \sqrt{1-\frac{L^2}{r^2}} d\tau dr + \frac{L^2}{r^2} dr^2 + r^2 d\omega^2. 
\end{equation}
This form of the metric resembles the Fefferman--Graham gauge on spatial hypersurfaces and the Bondi gauge in its time components. The transformation is valid for $r>L$. At $r=L$, the metric becomes Minkowski and we can attach a standard Minkowski slice along this timelike boundary in the interior as shown in Fig.~\ref{fig:example_c}. 
\end{example}

In the previous section, we distinguished hyperboloidal foliations from asymptotically null ones. From the point of view of numerical simulations, an important difference between these two families is the behavior of the outgoing characteristics speeds in compactifying coordinates. While the speeds are finite with hyperboloidal compactification, they are unbounded with asymptotically null compactification. Therefore, asymptotically null foliations cannot be used with explicit time integration schemes. They may, however, be useful in connection with implicit schemes \cite{dasilva2023hyperboloidaldiscontinuoustimesymmetricnumerical, markakis2023symmetric} or frequency domain computations \cite{macedo2022hyperboloidal, leather2024gravitational}. Below are three examples of height functions for stationary, asymptotically null slicings.

\begin{example}
    The mean extrinsic curvature is unbounded in asymptotically null foliations. Using this observation, we can modify the constant term in \eqref{eqn:gowdy_tau} such that the mean extrinsic curvature blows up,
\be
\label{eqn:hd}
-h_d(r) = \sqrt{\frac{1}{1+r}+r^2}, \qquad -H_d(r)= \frac{2r - \frac{1}{(1+r)^2}}{2\sqrt{\frac{1}{1+r}+r^2}}.
\ee
\end{example}
    
\begin{example}
Another approach is to modify the difference of the height function from the null case. Instead of a $1/r$-term in \eqref{eqn:ha}, we impose a faster decay such as,
\be
\label{eqn:he}
-h_e(r) = r + e^{-r}, \qquad -H_e(r)= 1 - e^{-r}, 
\ee
\end{example}

\begin{example}
The final example will play a role when we discuss de Sitter spacetimes in Sect.~\ref{sec:desitter},
\be
\label{eqn:hf}
-h_f(r) = \ln(\cosh r), \qquad -H_f(r)= \tanh r.
\ee
\end{example}

These three examples are plotted in Figs.~\ref{fig:example_d}, \ref{fig:example_e}, and \ref{fig:example_f}. Asymptotically null hypersurfaces intersect future null infinity at 45 degrees to the horizontal. However, the causal nature of the height functions depends on the large-scale structure of the spacetime. Asymptotically null foliations near null infinity in asymptotically flat spacetimes become horizon-penetrating near the event horizon or the cosmological horizon in combination with exponential mappings. In particular, the examples \eqref{eqn:he} and \eqref{eqn:hf} are horizon-penetrating in exponential compactification \cite{bizon_toy_2020}, discussed in Sec.~\ref{sec:spatial_compactifications}, 

The specific choice of the height function is more art than science. Some options are elegant and simplify calculations. Others provide flexibility but are messy (see \cite{macedo2020hyperboloidal} for an overview). The particular choice depends on the problem at hand, the background, and the compactification one prefers, which we discuss next. 

\subsection{Spatial compactification}\label{sec:spatial_compactifications}
A natural choice for a radial coordinate in a spherically symmetric spacetime is the areal coordinate. No such natural choice exists in the conformal completion. There are infinitely many functions that one can use for compactification. Mappings from unbounded domains to bounded domains for wave equations have been studied by Grosch and Orszag in 1977 \cite{GroschOrszag77} who distinguish algebraic and exponential mappings (see also \cite{boyd2001chebyshev,shen_spectral_2011}). In this section, we list common examples using this distinction.

A compactification involves a mapping $\rho=g(r)$ from an unbounded domain $r\in [0,\infty)$ to the bounded domain $\rho\in [0,S)$ such that
\begin{align}
& g(0)=0, \ g(\infty) = S, \nonumber \\
& \frac{dg}{dr} =: G(\rho) \geq 0, \quad G(S) = 0. \label{eqn:space_trafo}
\end{align}
The constant $S$ corresponds to infinity in the radial coordinate. In the examples, we take its value as $S=1$ or $S=\pi/2$ but the expressions can be trivially generalized to arbitrary values by a simple scaling of the compactified coordinate. Such scalings can be used to tune the spacing of points near the boundaries or change the domain size from $S$ to any other value \cite{shen_spectral_2011, Liu_Pynn_2016, macedo2022hyperboloidal}.

\newcommand{\resetexample}{\setcounter{example}{0}}
\renewcommand{\theexample}{\arabic{example}}
\resetexample
\begin{example}[Algebraic mappings]
\begin{align} 
& r = \frac{\rho}{1-\rho}, \quad && \rho = \frac{r}{1+r},   \quad && G = (1-\rho)^2. \label{eqn:linear_comp}\\
& r = \frac{2\rho}{1-\rho^2}, \quad &&\rho = \frac{-1 + \sqrt{1+r^2}}{r},   \quad &&G = \frac{(1-\rho^2)^2}{2 (1+\rho^2)}, \label{eqn:poincare}\\
& r = \tan \rho, \quad && \rho = \arctan r, \quad && G = \cos^2\rho. \label{eqn:trig}
\end{align}
\end{example}

\begin{example}[Exponential mappings]
\begin{align}
& r = \arctanh\ \rho = \frac{1}{2} \ln \frac{1+\rho}{1-\rho}, \quad && \rho = \tanh \left(r\right), \quad && G(\rho) = (1-\rho^2).\label{eqn:log}\\
& r = -\ln (1-\rho), \quad && \rho = 1-e^{-r}, \quad && G = 1-\rho. \label{eqn:exp}
\end{align}
\end{example}

Some of the examples above can be used to compress a domain that is unbounded in both directions to compactify both ends of the domain $r\in(-\infty,\infty)$ (\eqref{eqn:poincare}, \eqref{eqn:trig}, and \eqref{eqn:log}). This flexibility is relevant when compactifying the tortoise coordinate in black-hole spacetimes \cite{bernuzzi2012horizon}. If there are certain domains that require higher resolution, one can combine compactification with analytic mesh refinement \cite{macedo2022hyperboloidal}. The freedom in the spatial compactification can also be used to restrict the compactification to an annular domain called the hyperboloidal layer \cite{zenginouglu2011hyperboloidal, bernuzzi2011binary}. For such applications, the hyperboloidal time transformation must be combined with spatial compactification into hyperboloidal compactification. 

\subsection{Hyperboloidal compactification in Minkowski spacetime}

Hyperboloidal compactification \cite{zenginoglu_hyperboloidal_2008} combines the time transformation \eqref{eqn:time_trafo} with a spatial compactification \eqref{eqn:space_trafo} and, if needed, a conformal rescaling proportional to the compress function $G$.

The allowed combination of the time transformation and spatial compactification depends on the background spacetime. In high-dimensional Minkowski spacetime, only the algebraic mappings can be used with hyperboloidal transformation. The reason is the rate at which spheres grow with radius. To see this, we write the time-transformed Minkowski metric \eqref{eqn:theta_metric} in compactifying coordinates,
\begin{equation}\label{eqn:theta_rho_metric} ds^2 = \frac{1}{G} \left( -G \, d\tau^2 + 2 H \, d\tau d\rho + \frac{1-H^2}{G} d\rho^2 + G r^2 d\omega^2 \right).\end{equation}
The regularity of the conformal metric  $d\bar{s}^2 =G ds^2$ at the conformal boundary ($G=0$) demands that the metric coefficients $(1-H^2)/G$ and $G r^2$ are finite and non-vanishing. The term $G r^2$ vanishes at the conformal boundary for exponential mappings but has a finite value for algebraic mappings. Exponential mappings can be used in combination with the hyperboloidal transformation near cosmological horizons (Sect.~\ref{sec:desitter}).

Hyperboloidal compactification is a large family of transformations that lead to a stationary, globally regular foliation with asymptotically hyperbolic hypersurfaces. Within this large family, there are certain properties that stand out. For example, we can define a characteristic-preserving hyperboloidal compactification by maintaining the coordinate expression of the characteristics,
\[ \tau \pm \rho = t \pm r. \]
The characteristic-preserving property can be imposed to define either the time transformation or the compactification. It can only be satisfied in one direction (in- or outgoing). This method of constructing suitable hyperboloidal coordinates has proven useful in many numerical calculations \cite{zenginouglu2011hyperboloidal, bernuzzi2011binary, hilditch2018evolution}.     

The other family we encountered are the spatially flat hyperboloidal compactifications. In the case of Minkowski spacetime, the spatial flatness requirement implies for the boost and compress functions 
$1-H^2 = G$.
For example, combining the time function \eqref{eqn:gowdy_tau} with the algebraic compactification \eqref{eqn:poincare} scaled with $L$ gives Moncrief's spatially flat metric \eqref{eqn:mink_spatially_flat}. The spatial hypersurfaces give the Poincaré disk model for hyperbolic geometry. The off-diagonal shift term represents the outward flow of space, and the time slices of the conformal metric are spatially flat. This structure is similar to the Schwarzschild geometry in Painlevé-Gullstrand coordinates \cite{martel2001regular, hamilton2008river} or the de Sitter geometry in regularized static coordinates \cite{parikh_new_2002}. We can view stationary, horizon-penetrating coordinates as hyperboloidal compactification with exponential mappings. The similarities between these these hypersurfaces across event horizons and null infinity are captured in the notion of null-transverse hypersurfaces described in the next section. 

\section{Null-transverse hypersurfaces}\label{sec:4}
In Sec.~\ref{sec:2}, we discussed horizon-penetrating time surfaces that are transverse to the Schwarzschild event horizon. A regular, stationary time surface across the event horizon must necessarily avoid the bifurcation sphere because the timelike Killing field vanishes there. The two prominent examples are the spatially flat GP coordinates \eqref{eqn:gp_trafo} and the characteristic-preserving EF coordinates \eqref{eqn:ef_trafo}. Penrose coordinates are horizon-penetrating near the black hole and hyperboloidal near infinity but not stationary.

In Sec.~\ref{sec:3}, we discussed hyperboloidal time surfaces that are transverse to the Minkowski null infinity. A regular, stationary time surface across infinity must necessarily avoid spatial infinity because the timelike Killing field vanishes there. Generalizing the hyperboloid slicing \eqref{eqn:gowdy_tau} using a height-function, we discussed various examples of globally regular, stationary, hyperboloidal foliations.

Clearly, horizon-penetrating and hyperboloidal surfaces share certain properties. 
The difference between horizon-penetrating and hyperboloidal is, in the asymptotically flat case, that hyperboloidal coordinates approach null infinity and admit an algebraic compactification, whereas horizon-penetrating coordinates cross a finite horizon and admit an exponential compactification\footnote{The exponential compactification here refers to the relationship between the tortoise coordinate that is unbounded near the horizon, and the areal coordinate that has a finite value.}. Their common feature is that they cross a null horizon transversally. This null horizon can be a black hole, a cosmological horizon, or null infinity. The different cases can be captured with the following definition.

\begin{definition}
    Let \(\mathcal{N} \subset \mathcal{M}\) be a smooth null horizon, such as the event horizon of a black hole \(\mathcal{H}^\pm\), the cosmological horizon in an asymptotically de Sitter spacetime \(\mathcal{H}_c^\pm\), or null infinity in an asymptotically flat spacetime \(\scri^\pm\). A smooth, spacelike hypersurface \(\Sigma \subset \mathcal{M}\) is called \textbf{null-transverse} if:
\begin{enumerate}
    \item \(\Sigma\) intersects \(\mathcal{N}\) \textit{transversely}, meaning that for every point \(p \in \Sigma \cap \mathcal{N}\), $T_p \Sigma + T_p \mathcal{N} = T_p \mathcal{M}$. The intersection forms a codimension-2 submanifold, $\Sigma \cap \mathcal{N}$, which is typically diffeomorphic to \(S^{n-1}\) (e.g., a 2-sphere in 4-dimensional spacetimes),
    \item \(\Sigma\) is everywhere spacelike, i.e., the induced metric on \(\Sigma\) is Riemannian,
\end{enumerate}
\end{definition}
In short, a null-transverse spacelike hypersurface crosses a given null horizon smoothly without becoming singular or degenerate. In a stationary spacetime, we would like the null-transverse foliation to be also stationary. The stationary  GP \eqref{eqn:gp_trafo} and EF \eqref{eqn:ef_trafo} coordinates in Schwarzschild spacetime are null-transverse across the event horizon. The stationary hyperboloid slicing and the examples \eqref{eqn:ha}, \eqref{eqn:hb}, and \eqref{eqn:hc} in Minkowski spacetime are null-transverse across null infinity. Next, we discuss the stationary time transformation for a spherically symmetric metric to construct null-transverse foliations in de Sitter spacetime. In Sec.~\ref{sec:5}, we combine these ideas in the notion of a bridge that connects null horizons with a null-transverse hypersurface.

\subsection{Null-transverse surfaces in spherical symmetry}
We write a static, spherically symmetric metric in adapted coordinates as follows:
\begin{equation}\label{eq:metric} 
    ds^2 = - f(r) dt^2 + \frac{1}{f(r)} dr^2 + r^2 d\omega^2, 
\end{equation}
We assume that the metric coefficient $f(r)$ is a smooth function with positive, real zeros $r_i$. For $f>0$, the metric is static with timelike Killing field $\partial_t$. The Killing field becomes spacelike whenever $f<0$, which happens beyond null horizons. The coordinate patch in which the timelike Killing field is manifest does not necessarily cover the entire spacetime. In this paper, we are interested in the patch in which the metric can be written in the above form.

Many textbook examples of spherically symmetric spacetimes can be written in this form. As discussed previously, we get Minkowski with $f=1$, Schwarzschild with $f=1-\frac{2M}{r}$, where $M$ is the mass of the black hole, de Sitter with $ f = 1-\frac{r^2}{L^2}$, where $L$ is the curvature radius, Schwarzschild--de Sitter (Kottler) with $f=1-\frac{2M}{r} -\frac{r^2}{L^2}$, and Reissner-Nordström with $f = 1-\frac{2M}{r} + \frac{Q^2}{r^2}$, where $Q$ is charge. The metric as written in \eqref{eq:metric} is singular at the roots of $f$. For Schwarzschild spacetime, this is the well-known Schwarzschild radius at $r_b=2M$, discussed extensively in Sec.~\ref{sec:2}. In de Sitter spacetime, this is the cosmological horizon at $r_c=L$. Compactification shows that the coordinates are singular also at infinity. The black-hole horizon, the cosmological horizon, and null infinity are very different physically, but share causal features. 

\begin{figure}[h]%
    \centering
    \includegraphics[width=0.49\textwidth]{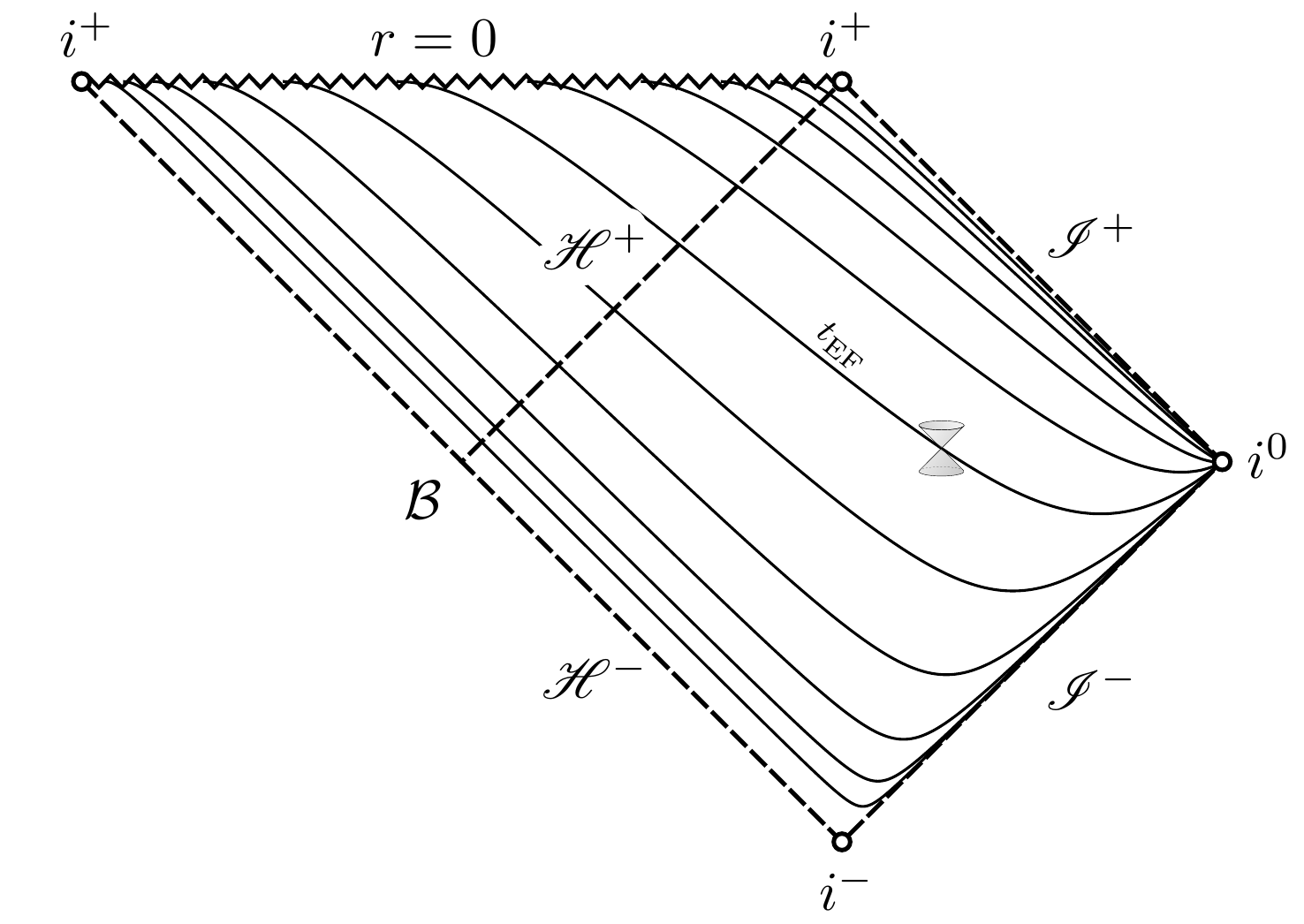} \hspace{-1cm}
    \includegraphics[width=0.49\textwidth]{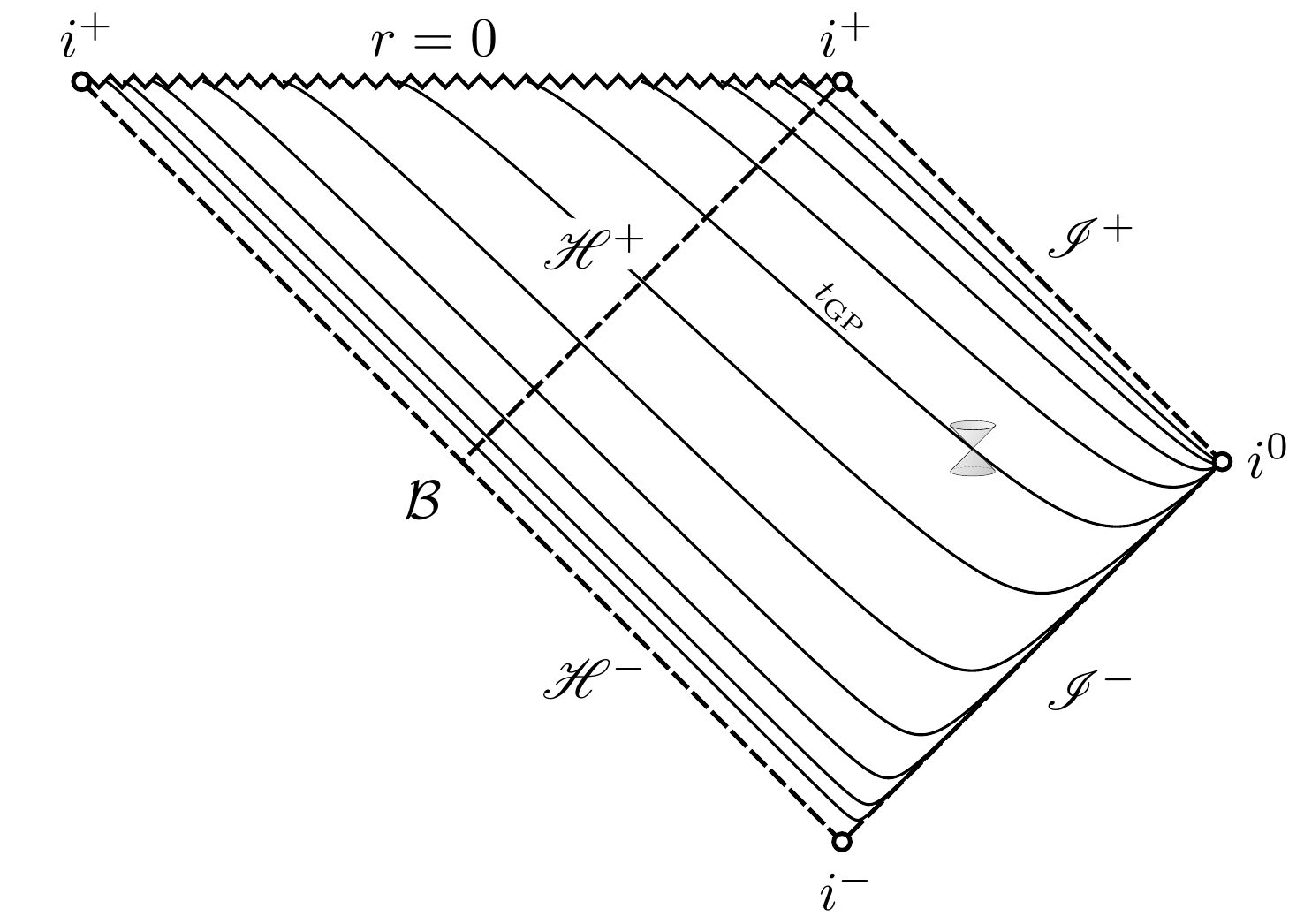}    
    \caption{Stationary, regular foliation of the Schwarzschild event horizon with null-transverse hypersurfaces of the GP \eqref{eqn:gp_trafo} and EF \eqref{eqn:ef_trafo} time coordinates.} \label{fig:ef_gp}
\end{figure}

As in the previous sections, we resolve the coordinate singularity by switching to a regular time coordinate that maintains stationarity. In particular, we require that the new time coordinate, $\tau$, satisfies $\partial_\tau = \partial_t$, possibly up to a constant rescaling. Therefore, we use a height function to define the new time coordinate as in \eqref{eqn:height_function}.  The transformed metric becomes
\begin{equation} \label{eq:metric_tau}
    ds^2 = - f d\tau^2 + 2 f H \ d\tau dr + \frac{1-f^2 H^2}{f} dr^2 + r^2 d\omega^2, 
\end{equation}
The metric is simply a slightly more general version of the Minkowski case discussed in \eqref{eqn:theta_metric}. The regularity requirement for the resulting metric imposes conditions on the boost function similar to the hyperboloidal case \eqref{eqn:time_trafo}. Consider the domain $(r_i,r_o)$ which represents the static patch where the metric takes the above form with $f(r_i)=0$ and $r_o>r_i$ some outer radius. We require
\begin{equation}\label{eqn:time_trafo_f} |f H|<1 \quad \forall r\in (r_i,r_o) \quad \& \quad |f H| = 1 - \mathcal{O}(f)\  \mathrm{as}\ r\to r_i, \end{equation}
If the boost function satisfies these properties, we obtain a stationary, regular foliation across the null horizon. A singular boost is required at the roots of $f$ to regularize the static time with a singularity precisely of the order dictated by $f$. Both GP \eqref{eqn:gp_trafo} and EF \eqref{eqn:ef_trafo} coordinates satisfy these properties near $r_b=2M$ with
\[ f H_\gp = \pm \sqrt{\frac{2M}{r}}, \qquad f H_\ef = \pm \frac{2M}{r}. \]
These stationary foliations with future-directed hypersurfaces are depicted on a Penrose diagram in Fig.~\ref{fig:ef_gp}. The leaves of the foliation are null-transverse at the future black hole horizon. Comparing with Fig.~\ref{fig:maximal}, we see that the time slices for $t_\ef$ and $t_\gp$ avoid the bifurcation sphere, $\mathcal{B}$, and foliate the future event horizon $\mathcal{H}^+$. However, the slices still intersect at spatial infinity. Next, we discuss null-transverse coordinates in de Sitter spacetime across the cosmological horizon.

\vspace{-7mm}

\subsection{Stationary, null-transverse time for the de Sitter metric}\label{sec:desitter}

The path to the resolution of the Schwarzschild coordinate singularity reviewed in Sect.~\ref{sec:2} was historically paved by a deeper understanding of de Sitter's cosmological solution by Lemaître \cite{lemaitre1933univers}. The de Sitter metric has a positive cosmological constant $\Lambda$, which defines a cosmological length scale $L$ with $\Lambda = 3/L^2$. The metric can be written in the form \eqref{eq:metric} with
\begin{equation}\label{eqn:f_desitter} f = 1-\frac{r^2}{L^2}.\end{equation}

\begin{figure}[h]%
    \centering
    \includegraphics[width=0.37\textwidth]{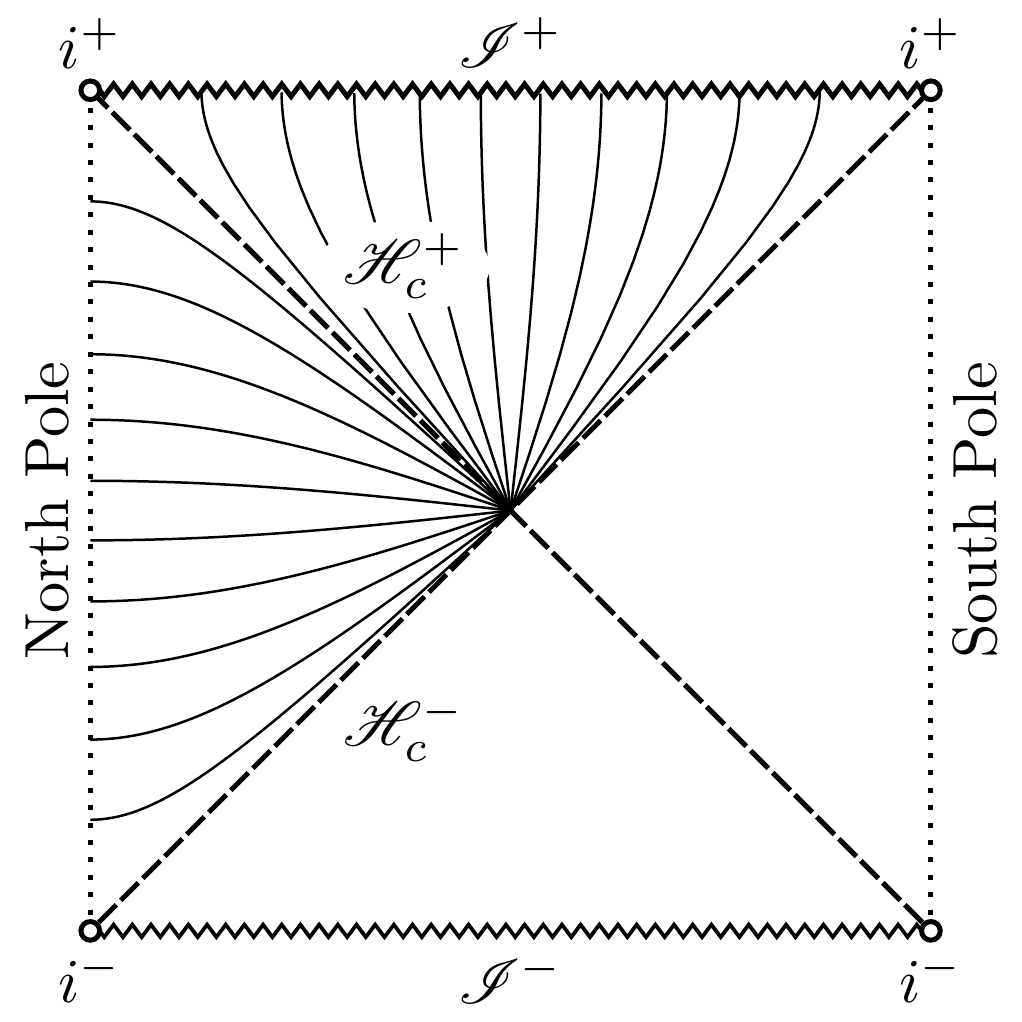}
    \caption{Conformal diagram of de Sitter spacetime. The solid lines are level sets of the time coordinate $t$.} 
    \label{fig:desitter}
\end{figure}

We can view de Sitter spacetime as a black-hole solution inside out. Both the Schwarzschild radius and the de Sitter radius are coordinate singularities. A particle dropped from a finite radius takes an infinite static time to reach the horizons, but a finite proper time as measured by a comoving observer. As a consequence, these horizons are neither singular nor inaccessible \cite{israel1987dark}. It is the static time that is singular due to the intersection of its level sets as shown in the conformal diagram of de Sitter spacetime in Fig.~\ref{fig:desitter}. 

Even though the nature of the coordinate singularity was already understood by Lemaître, regular and stationary time coordinates across the de Sitter horizon were constructed relatively recently as we review next.

\subsubsection{The minimum height function}
We inspect the tortoise coordinate to find the leading-order height function that ensures a regular, null-transverse foliation,
\be\label{eqn:tortoise_desitter}
\frac{dr}{f} = dr_\ast \quad \implies \quad  
 r_\ast =  L \arctanh\frac{r}{L} = \frac{1}{2\kappa} \left[\ln \left(1+\frac{r}{L}\right) - \ln\left(1-\frac{r}{L}\right)\right].
\ee
The surface gravity of the de Sitter horizon is $\kappa = 1/L$. We would like the foliation to behave like an outgoing characteristic, $u=t-r_\ast$, while remaining spacelike. We can satisfy $h\sim -r_\ast$, by taking the leading-order term in the expression for the tortoise function with a negative sign,
\be\label{eqref:minimum_dS} h =  \frac{1}{2\kappa} \ln\left(1-\frac{r}{L}\right), \quad f H = - \frac{1}{2} \left(1+ \frac{r}{L} \right).\ee
The transformed metric is regular at the cosmological horizon.

\subsubsection{Spatially flat (Parikh)}
\begin{figure}[h]%
    \centering
    \includegraphics[width=0.24\textwidth]{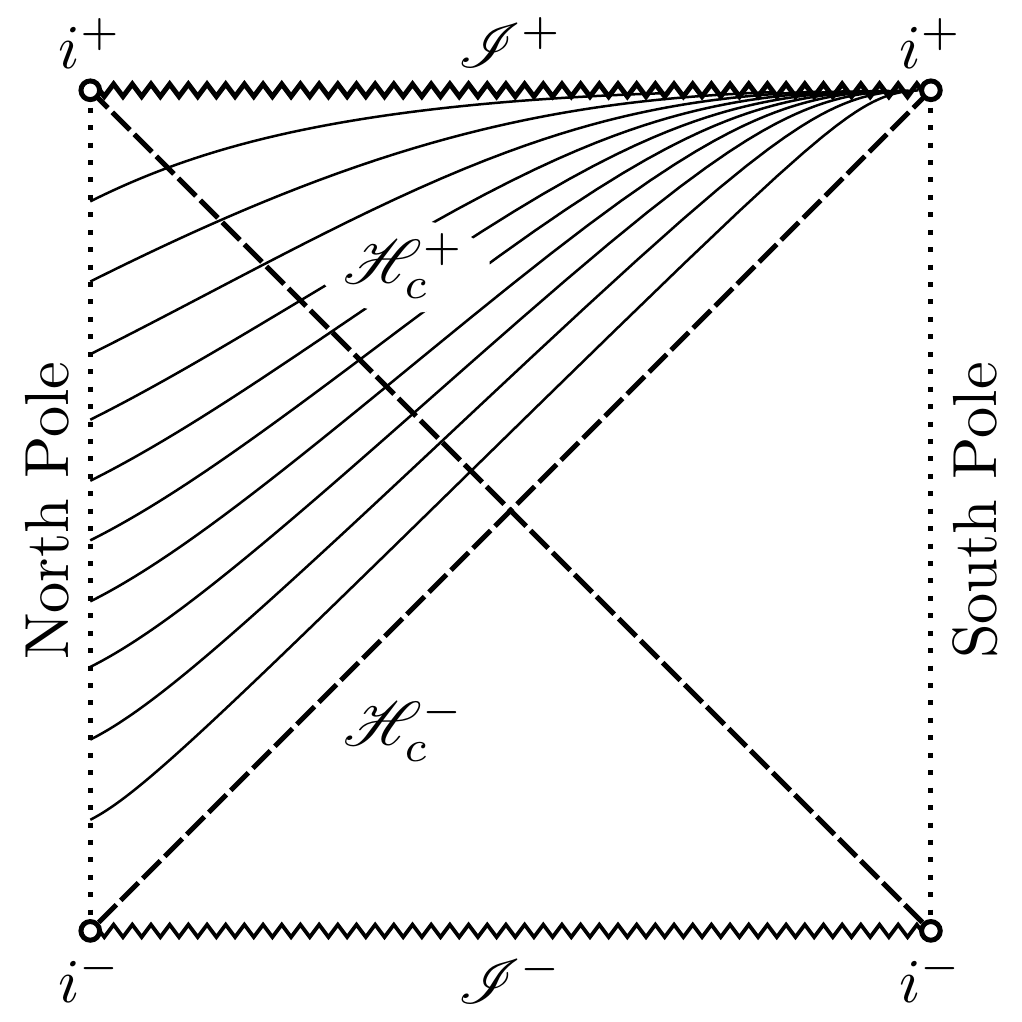}
    \includegraphics[width=0.24\textwidth]{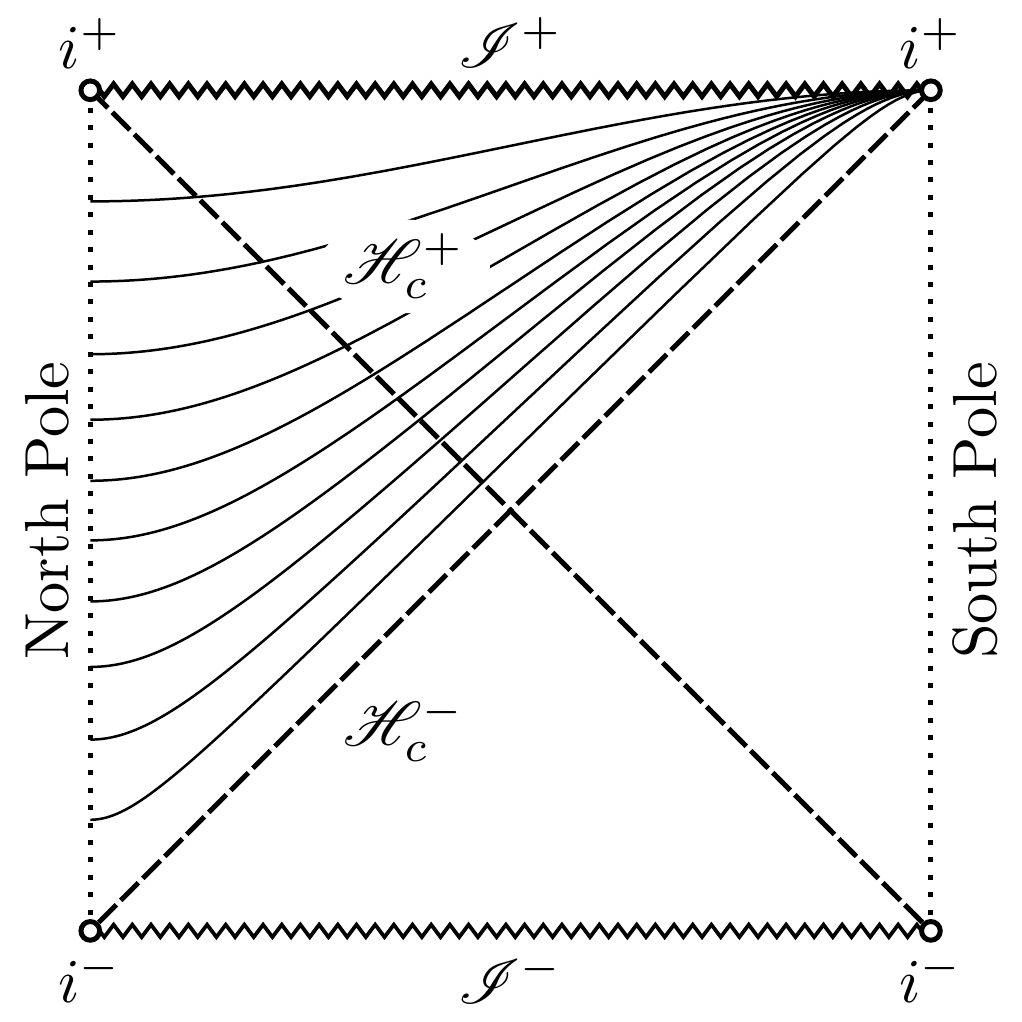}
    \includegraphics[width=0.24\textwidth]{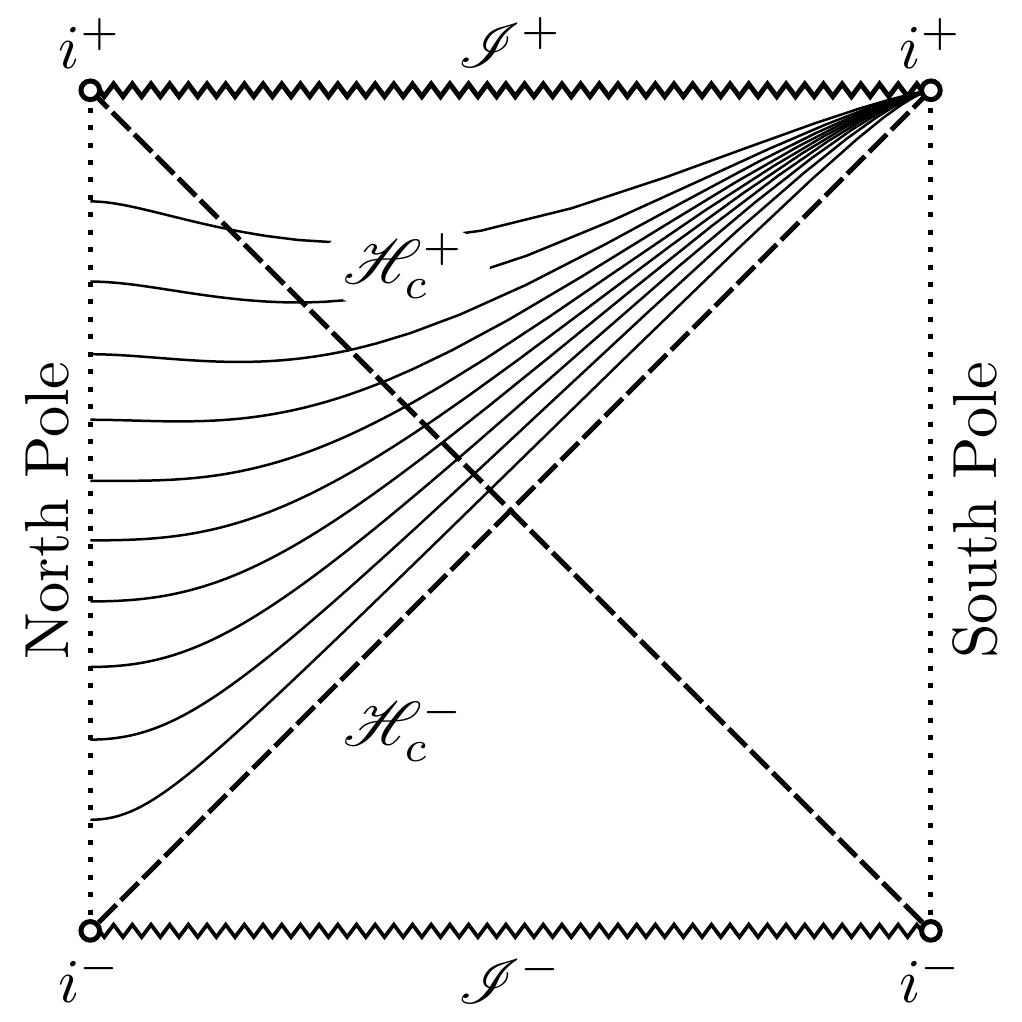}
    \includegraphics[width=0.24\textwidth]{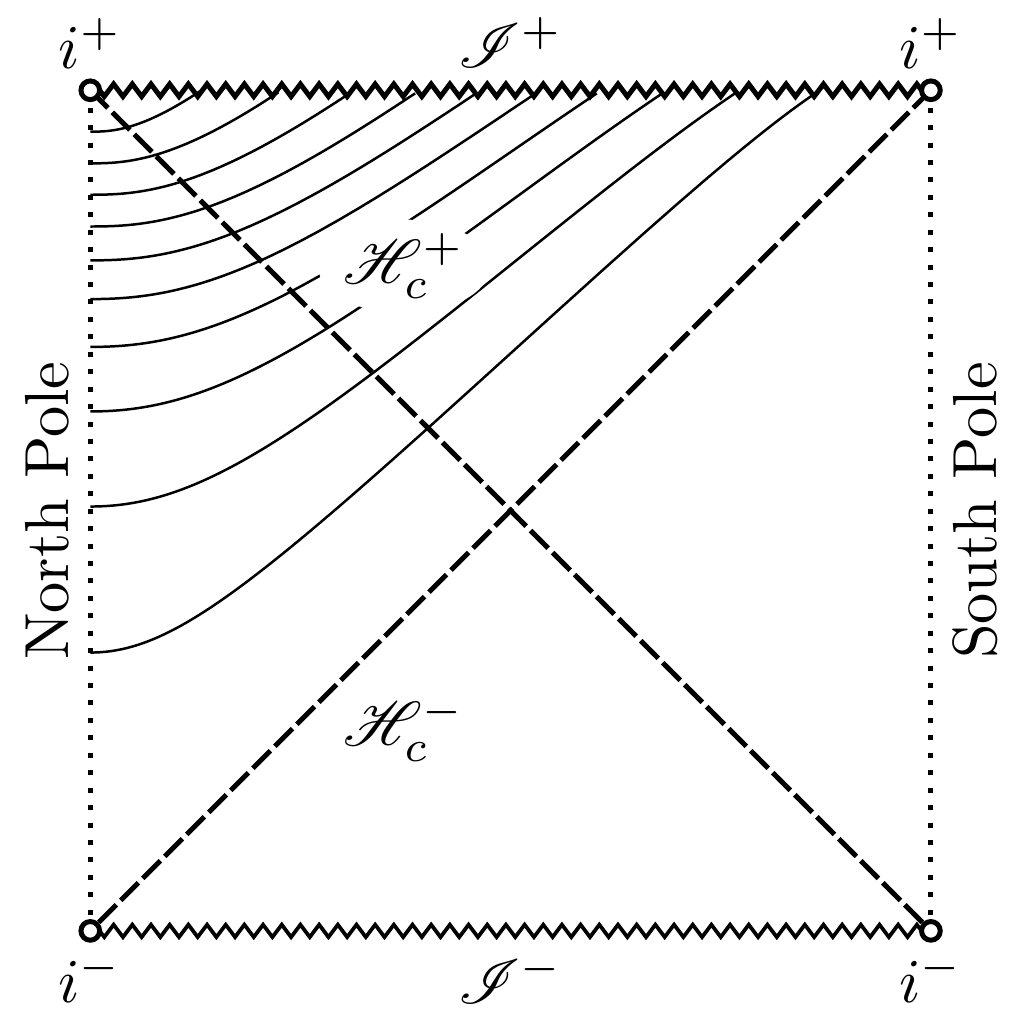}
    \caption{Conformal diagrams for minimum \eqref{eqref:minimum_dS}, spatially flat \eqref{eqn:spatially_flat_desitter}, characteristic-preserving \eqref{eqn:kerrschild}, and Misner \eqref{eqn:misner} time slices in de Sitter spacetime.}
    \label{fig:desitter_examples}
\end{figure}

The PG form of de Sitter with spatially flat hypersurfaces was constructed by Parikh in \cite{parikh_new_2002} (see also \cite{bizon_toy_2020, hintz2021quasinormal}). We set $\frac{1-f^2 H^2}{f} = 1$ with $f$ given by \eqref{eqn:f_desitter} to get
\be\label{eqn:spatially_flat_desitter} f H = \pm \frac{r}{L} \ \implies \ h = \pm \frac{L}{2}\ln f = \pm \frac{1}{2\kappa} \left[ \ln\left( 1 + \frac{r}{L}\right) + \ln\left( 1-\frac{r}{L}\right) \right]. \ee
The positive sign corresponds to a future horizon-penetrating foliation. We have,
\be\label{eqn:dS_Parikh} ds^2 = -f d\tau^2 \mp \frac{2r}{L} d\tau dr + dr^2 + r^2 d\omega^2\,.\ee
The metric is regular at the horizon, stationary, and spatially flat. The conformal diagram of the slices is given on the second to the left panel of Fig.~\ref{fig:desitter_examples}.

\subsubsection{Characteristic-preserving (Kerr-Schild)}
Similar to the construction of Eddington-Finkelstein coordinates, we demand that the coordinate expression of the characteristic approaching the horizon is preserved. Note that the areal coordinate $r$ can be considered as the compactification of the tortoise coordinate $r_\ast$ where $f$ in \eqref{eqn:tortoise_desitter} plays a similar role as $G$ in \eqref{eqn:space_trafo}. When the horizon is at a finite distance, the spatial compactification is exponential. Imposing the characteristic-preserving condition, we get
\be \label{eqn:kerrschild}\tau - r = t - r_\ast \quad \implies \quad h = r - r_\ast \quad  \implies \quad fH = -\frac{r^2}{L^2}. \ee
The metric takes the Kerr-Schild form,
\begin{align} \label{eq:dS_KS}
ds^2 &= -f d\tau^2 - \frac{2r^2}{L^2} d\tau dr + 
\left( 1+\frac{r^2}{L^2} \right) dr^2 + r^2 d\omega^2. \\ 
&= - d\tau^2 + dr^2 + r^2 d\omega^2 + \frac{r^2}{L^2} (d\tau-dr)^2. \nonumber
\end{align}    

The three foliations we discussed are stationary by construction, spacelike, non-intersecting, and horizon-penetrating. The similarity between hyperboloidal slices near null infinity in Fig.~\ref{fig:mink_penrose} and horizon-penetrating slices near the cosmological horizon in the static patch of de Sitter in Fig.~\ref{fig:desitter_examples} demonstrates common causal features between these two cases. However, there is an important distinction. The region beyond null infinity in asymptotically flat spacetimes is unphysical whereas the region beyond the cosmological horizon may be relevant for certain computations. In particular, the behavior of the coordinates in this region is relevant for the flat limit.

The conformal diagrams show that the time slices intersect beyond the cosmological horizon. Thinking of de Sitter space as an inside out black hole, this intersection is similar to what happens with trumpet slices in Schwarzschild spacetime near the black hole singularity \cite{Hannam_2008, Dennison_2014, vano2024conformal}. In the flat limit as $L\to\infty$, the metrics \eqref{eqn:dS_Parikh} and \eqref{eqn:kerrschild} approach the flat metric in standard coordinates and the intersection of time slices happens at spatial infinity. To obtain a regular flat limit, Misner combined the Kerr-Schild transformation with a flat, hyperboloid term to open up the time slices beyond the cosmological horizon  \cite{misner_excising_2006, misner_over_2006, misner_hyperboloidal_2006}.

\subsubsection{Misner}

Misner used the Kerr-Schild form of the de Sitter metric as the starting point for a model problem in hyperboloidal compactification \cite{misner_excising_2006, misner_over_2006, misner_hyperboloidal_2006}. Modifying the characteristic-preserving height function \eqref{eqn:kerrschild} with a hyperboloid transformation in $r$ \eqref{eqn:gowdy_tau} resolves the intersection of the time slices at infinity beyond the cosmological horizon. This transformation does not change the regularity of the coordinates near the cosmological horizon. Its purpose is to obtain a regular flat limit. We define
\be\label{eqn:misner}
h = r - r_\ast - \sqrt{s^2+r^2}.
\ee
The constant $s$ determines how fast the slicing resembles an outgoing null slicing approaching $\scri^+$. The time slices with $s=1$ are plotted on the right panel of Fig.~\ref{fig:desitter_examples}. 

Misner combined this transformation with an algebraic spatial compactification of the form \eqref{eqn:poincare} to discuss Maxwell equations \cite{misner_excising_2006}. This compactification is not needed if one is only interested in the behavior of the fields up to the cosmological horizon because the horizon is at a finite distance. The advantage of this coordinate system is rather in its flat limit: as $L\to\infty$, one recovers the hyperboloid slicing of Minkowski spacetime. This modification allows one then to use a small cosmological constant to simulate fields in de Sitter spacetime.

Another option that makes use of the hyperboloid slicing but is adapted to the causal structure of de Sitter space near the horizon is to use the tortoise coordinate: $h = -\sqrt{1+r_\ast^2}$. Such a slicing can be used in other spacetimes as well (compare \eqref{eqn:sa_hypal}).

    

\subsection{Hyperboloidal compactification in spherical symmetry}
In the case of null infinity, the construction of null-transverse surfaces includes spatial compactification because null infinity is infinitely far away (see \eqref{eqn:theta_rho_metric}). The situation is not qualitatively different from black-hole or cosmological horizons. Here, we give a summary of the general procedure in spherical symmetry. Hyperboloidal compactification introduces new coordinates $\tau$ and $\rho$ by
\[  \rho = g(r), \qquad \tau = t + h(r). \]
Defining, as before,
\[ \qquad G(\rho):=\frac{dg}{dr}, \qquad H(\rho):=\frac{dh}{dr} = G \frac{dh}{d\rho}, \]
the metric \eqref{eq:metric} becomes
\be\label{eqn:compactified_metric} ds^2 = \frac{1}{G} \left(-f G d\tau^2 + 2 f H d\tau d\rho + \frac{1 - (f H)^2}{f G} d\rho^2 + G r^2 d\omega^2\right). \ee
Suitable choices of $G$ and $H$ discussed in previous sections ensure that the metric above is regular across the roots of $f$ and $G$.

    

\section{Bridge foliations}\label{sec:5}

Gravitational wave astronomy connects the dynamics of strong gravitational fields near black holes with observations far away. Therefore, a regular time coordinate should connect the observer with the black hole(s). Until recently, black-hole perturbation theory was typically performed using Schwarzschild or Boyer-Lindquist time. These coordinates are not regular across the relevant null horizons modeling the black hole and the asymptotic region. In previous sections, we constructed null-transverse time surfaces near event horizons (Sec.~\ref{sec:2}), null infinity (Sec.~\ref{sec:3}), and the cosmological horizon (Sec.~\ref{sec:4}). In this section, we connect distinct horizon regions together in the notion of a bridge that is null-transverse at its ends.


\begin{definition}
Let \((\mathcal{M}, g)\) be a smooth, asymptotically flat or asymptotically de Sitter spacetime that contains at least one event horizon \(\mathcal{H}^+\). A \textbf{bridge} is a spacelike hypersurface \(\Sigma \subset \mathcal{M}\) that is null-transverse across all relevant null horizons—specifically, the event horizon(s) near the black hole(s) and null infinity or the cosmological horizon.
\end{definition}
We call a foliation consisting of bridges a bridge foliation. A stationary bridge foliation is time-translation invariant. The underlying concept is not new—such coordinates have been used in black-hole perturbation theory under various names, such as horizon-penetrating-hyperboloidally-compactified \cite{ripley2022computing} or simply hyperboloidal. We use the term bridge instead of hyperboloidal because the original definition of hyperboloidal surfaces includes spacelike slices across null infinity with asymptotically hyperbolic geometry—a condition that does not hold across finite horizons. In the next two sections, we discuss examples of stationary bridge foliations in the asymptotically flat case represented by Schwarzschild spacetime (Sec.~\ref{sec:ss_bridge}) and the asymptotically de Sitter case represented by Schwarzschild--de Sitter spacetime (Sec.~\ref{sec:sds_bridge}).

\subsection{Schwarzschild bridges}
\label{sec:ss_bridge}

Historically, Penrose time in Schwarzschild spacetime discussed in Sec.~\ref{sec:2_conformal} was the first bridge foliation \cite{penrose_zero_1965, carter1966complete}. The first stationary bridge foliation was given by constant mean curvature (CMC) surfaces \cite{brill_k_1980}. A bridge in Schwarzschild spacetime must satisfy the regularity conditions at each end of the domain of interest that we can read off from the tortoise coordinate \eqref{eqn:tortoise_k}. Near the event horizon at $r_b = 2M$, the leading-order behavior for a future-directed horizon-penetrating foliation is
\be\label{eqn:ss_horizon} h \sim +r_\ast \sim \frac{1}{2\kappa_b} \ln |f|, \quad \mathrm{as} \quad r\to r_b.\ee
Near null infinity, the leading-order behavior for a future-directed hyperboloidal foliation is
\be\label{eqn:ss_infinity} h \sim  - r_\ast \sim -r - \frac{1}{2\kappa_b} \ln r, \quad \mathrm{as} \quad r\to \infty. \ee
All examples discussed below satisfy these conditions.

\begin{example}[Minimal gauge]
Arguably, the most elegant construction of a stationary bridge is the \textit{minimal gauge} \cite{schinkel_initial_2014,PanossoMacedo:2018hab,PanossoMacedo:2023qzp}. It plays a central role in many of the applications of the hyperboloidal approach, such as the pseudospectrum of quasinormal modes \cite{jaramillo2021pseudospectrum, destounis2021pseudospectrum, cao2024pseudospectrum, Panosso_Macedo_2025}. As the name suggests, minimal gauge combines the minimum requirements for a bridge and therefore provides a good starting point to understand other examples. We simply add the two required behaviors \eqref{eqn:ss_horizon} and \eqref{eqn:ss_infinity} to obtain the minimal gauge slicing,
\be\label{eqn:minimal_gauge}
h = \frac{1}{2\kappa_b} \left(\ln |f| - \ln r\right) -r   \ \implies \ fH = -1 + \frac{2 r_b^2}{r^2}.
\ee
There are other ways to understand this gauge. For example, at the level of the boost function, we can impose the conditions on $fH$ derived from the regularity of the metric \eqref{eqn:compactified_metric} at the black hole horizon and at null infinity with an algebraic compactification. We require
\[ fH |_{r=r_b} = 1, \qquad fH|_{r\to\infty} \sim -1 + \frac{C}{r^2}, \]
with a constant $C$. This requirement is minimal in the sense that no subleading orders in $r$ are included for a regular, spacelike foliation. By evaluating the asymptotic form at the horizon $r=r_b$, we see that $C=2 r_b^2$, which gives the minimal gauge \eqref{eqn:minimal_gauge}. The original derivation in \cite{schinkel_initial_2014} uses ingoing coordinates. For a detailed discussion of the minimal gauge, including alternative constructions for different spacetimes, see \cite{PanossoMacedo:2023qzp}. 
\end{example}

\begin{figure}[h]%
    \centering
    \includegraphics[width=0.39\textwidth]{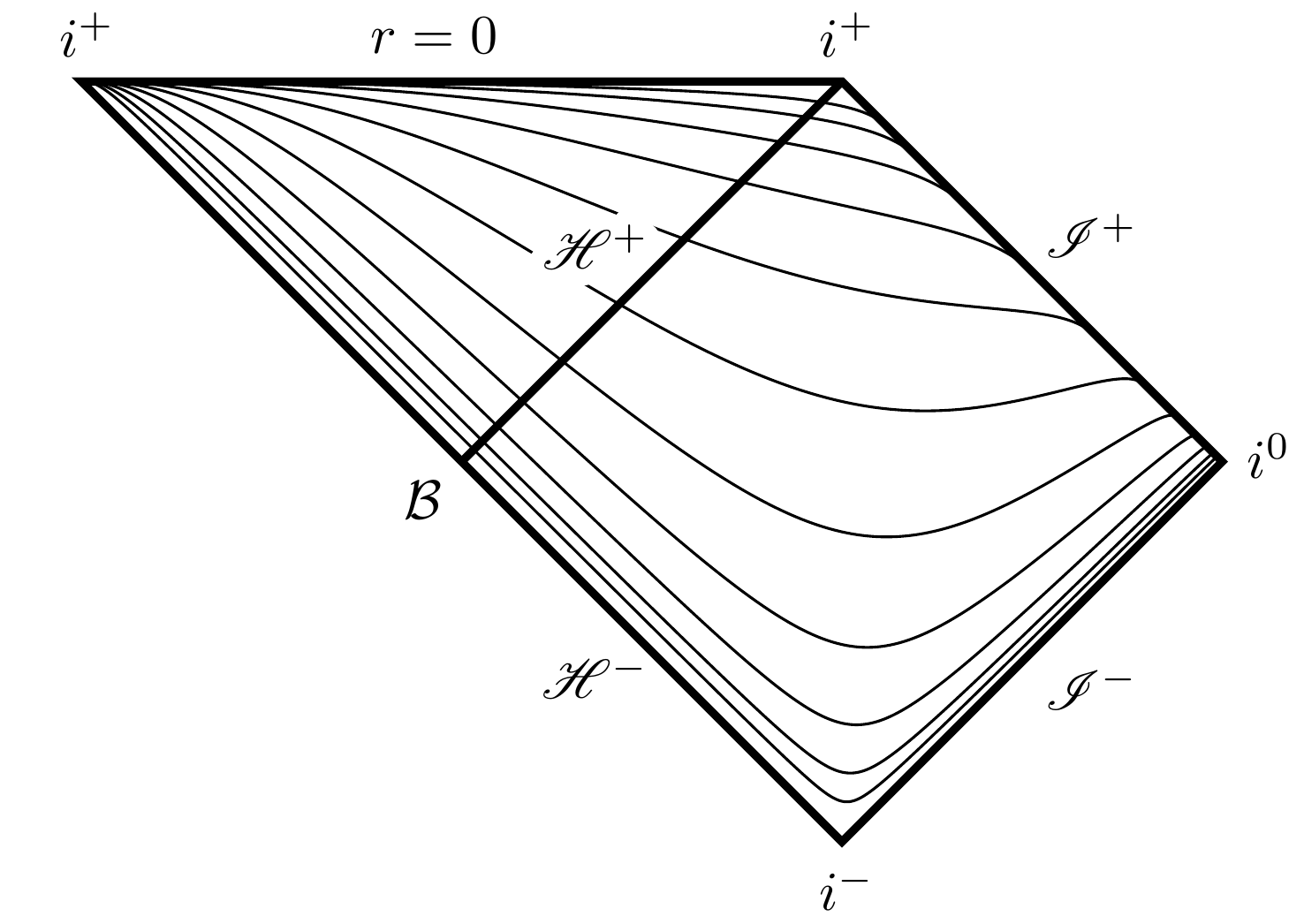} \hspace{-13mm} 
    \includegraphics[width=0.39\textwidth]{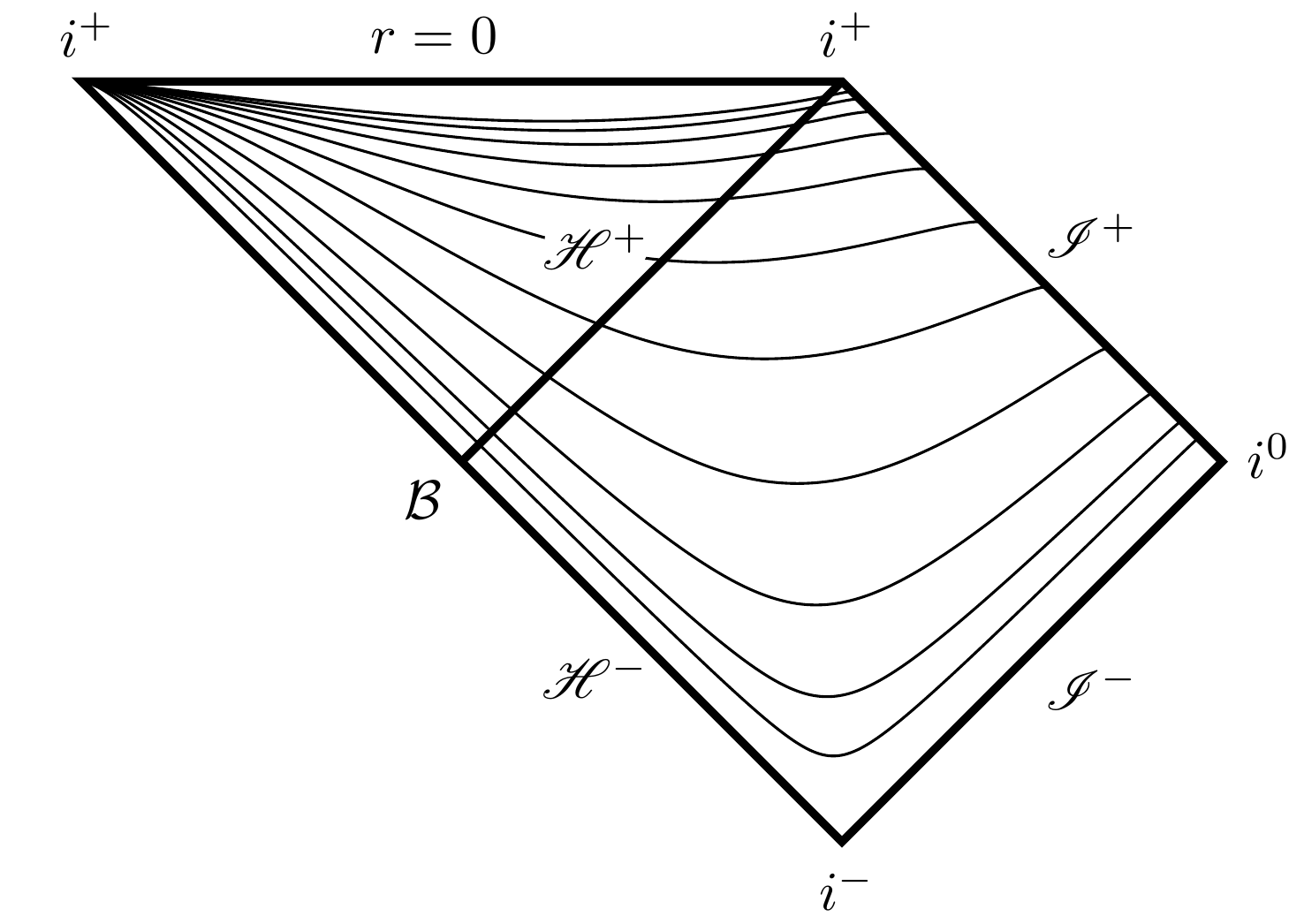}  \hspace{-13mm}  
    \includegraphics[width=0.39\textwidth]{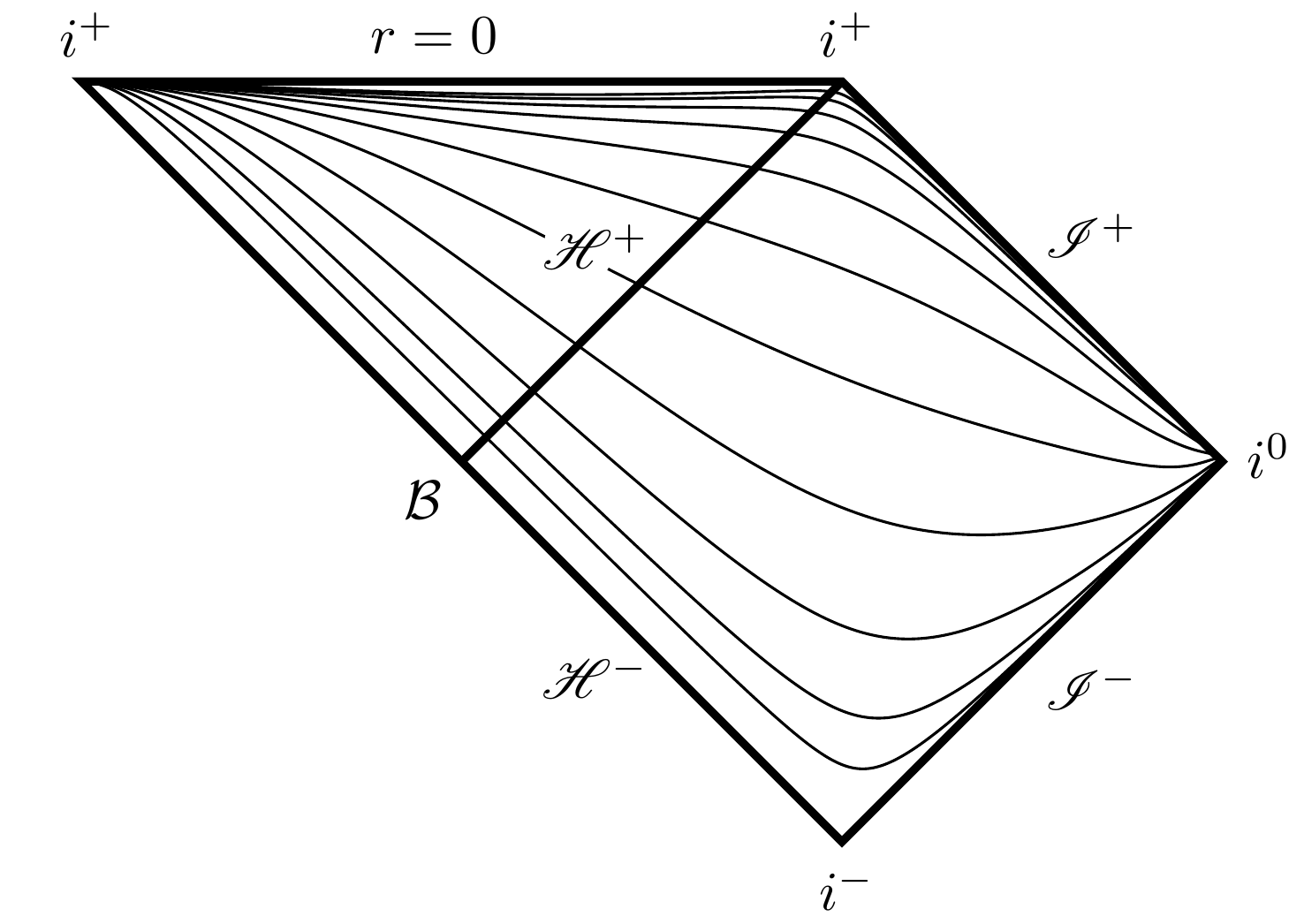}    
    \caption{Penrose diagrams for bridges in Schwarzschild spacetime connecting the future event horizon, $\mathcal{H}^+$, and future null infinity, $\scri^+$. On the left is the minimal gauge slicing \eqref{eqn:minimal_gauge}. The middle and the right diagrams are source-adapted \eqref{eqn:source_adapted} with $r_t=2.4 M$ and $r_t=8M$. As the turning point moves further out, the hypersurfaces become outgoing a narrower region closer to future null infinity.} \label{fig:ss_bridges}
\end{figure}

The sign of the term $fH$ determines whether the hypersurface is primarily ingoing or outgoing. Every future-directed bridge has at least one turning radius at which the hypersurface switches from an ingoing one to an outgoing one (see conformal diagrams in Fig.~\ref{fig:ss_bridges}). This turning radius is located at the root of $fH$. For the minimal gauge, the turning radius is at $r_t = \sqrt{2}r_b$. If there is a source of radiation farther away from this turning radius, the incoming radiation fights against the outgoing foliation to fall into the black hole, and gets blue-shifted until it passes the turning radius. In such cases, it may be beneficial to adjust the foliation such that the turning point is close to the source of the perturbations.

\begin{example}[Source-adapted bridge]
We construct a source-adapted bridge foliation by adding a higher-order term to the minimal gauge \eqref{eqn:minimal_gauge}. Denoting the event horizon as $r_h$ and the turning radius as $r_t$ with $r_t>r_h$, we set
\be\label{eqn:source_adapted} fH = (r_t^2-r^2) \left( \frac{r_b^4}{r^4 (r_t^2-r_b^2)} + \frac{r^2-r_b^2}{r^4+r_t^4}\right). \ee
This expression satisfies the conditions for a future-directed, bridge boost function. The corresponding slices are plotted on the middle and right panels of Fig.~\ref{fig:ss_bridges} with $r_t=2.4M$ and $r_t=8M$. For larger values of the turning point, the bridge becomes outgoing farther out corresponding to a narrower region close to future null infinity in the conformal diagram, which corresponds to the half-space interval, $r\in(r_t,\infty)$. 
\end{example}

\begin{example}[Source-adapted hyperboloids] The source-adapted bridge in \eqref{eqn:source_adapted} seems somewhat ad hoc. There are more elegant alternatives to modify the foliation to adapt to the location of a source. One such alternative is based on the off-centered hyperboloids \eqref{eqn:hb}. We obtain a source-adapted bridge foliation by replacing the areal coordinate with the tortoise coordinate,
\begin{equation}\label{eqn:sa_hypal}
    -h = \sqrt{L^2 + (r_\ast-r_t)^2}, \qquad -f H= \frac{r_\ast-r_t}{\sqrt{L^2 + (r_\ast-r_t)^2}}.
\end{equation}
\end{example}

Next, we mention a few more examples that have been used in calculations of black-hole perturbations in Kerr spacetimes but in their Schwarzschild limit of vanishing angular momentum.

\begin{example}[Rácz-Tóth]
Rácz and Tóth use a bridge foliation to study late-time tail decay rates in Kerr spacetime \cite{racz2011numerical, R_cz_2024}. They apply the spatial compactification \eqref{eqn:poincare} and extend Moncrief's hyperboloidal compactification for the flat case \cite{Moncrief00, FodorRacz04, fodor2008numerical, bizon2009universality} to the black-hole case. Translated to the areal coordinate, their height function reads,
\be h = 2M \left(\ln |f| - \ln r\right) - \sqrt{1+r^2} + 4 M \ln \left[\frac{2(-1+\sqrt{1+r^2})}{r}\right].
\ee
We recognize the flat-space hyperboloid \eqref{eqn:gowdy_tau} complemented with the logarithmic term necessary for the regularity of the coordinates for non-vanishing mass.
\end{example}

\begin{example}[HH]
Another bridge foliation for Kerr spacetime is based on the characteristic-preserving condition written in ingoing coordinates with the spatial compactification \eqref{eqn:linear_comp} \cite{harms2014new}. The coordinates are referred to as HH for horizon-penetrating and hyperboloidal. The height function written in the areal coordinate for the Schwarzschild case reads
\be h = 2M \left(\ln |f| - \ln r\right) - r + \frac{r}{1+r} + 4 M \ln \frac{r}{1+r}.
\ee
The HH bridge has a similar structure as the Rácz-Tóth bridge but different numerical properties in wave propagation problems \cite{harms2014new}. The hyperboloidal term is, up to a constant, the simple height function in Minkowski spacetime \eqref{eqn:ha}. 
\end{example}

\begin{example}[Mavrogiannis]
Bridges are also used in mathematical analysis of wave equations. A relatively simple foliation is given by Mavrogiannis in
\cite{mavrogiannis_quasilinear_2021, mavrogiannis2023morawetz} with the boost function
\be\label{eqn:mavro} -fH = \left(1-\frac{3M}{r}\right) \sqrt{1+\frac{6M}{r}}. \ee
The turning point of the Mavrogiannis bridge is at the photon sphere, $r_t=3M$. This makes sense because, inside the photon sphere, the inward pull of the black hole dominates for most null geodesics. The height function can be written as
\[ h = 2 M (\ln |f| - \ln r) - \sqrt{r(r+6M)} - 4 M \ln \left(1+\frac{2M}{r}+\sqrt{\frac{r+6M}{r}}\right). \]
\end{example}

\begin{example}[CMC] 
Historically, the first stationary bridge was the CMC foliation \cite{brill_k_1980} given via 
\[ -f H = \frac{J}{\sqrt{J^2 + f r^4}}, \qquad J = \frac{r^3}{L} - C, \]
where $L$ and $C$ are constants with $C>8 M^3/L$ and $L>0$ for a future hyperboloidal foliation. This foliation was used in the first numerical computations using bridges \cite{zengin2008hyperboloidal, zenginouglu2009gravitational, zenginouglu2010asymptotics, bizon_saddle-point_2010}. They have also been used to solve the hyperboloidal initial value problem for Einstein equations in a constrained formalism \cite{moncrief2009regularity, Rinne_2010, bardeen2011tetrad, rinne2013hyperboloidal, morales2017evolution}.
\end{example}

CMC slices have the same difficulty as characteristic slices: they are defined through a rigid, local condition. Their extension to other cases, such as the Kerr metric, is given only numerically \cite{schinkel2014axisymmetric}. In contrast, bridges are defined mainly through asymptotic conditions. The only local condition is that they are spacelike.The flexibility and adaptability of bridges is an important advantage for numerical work. To make the best use of this flexibility, a rigorous numerical analysis of their properties for solving hyperbolic equations should be performed. Eventually, we can expect that guidelines will be developed for choosing suitable bridges for different scenarios of interest.

\subsection{Schwarzschild--de Sitter bridges}
\label{sec:sds_bridge}

One of the motivations for including the terminology of null-transverse hypersurfaces and bridge foliations is to have a uniform language that covers the modeling of isolated systems and cosmological spacetimes. The term hyperboloidal is typically used in the context of asymptotically flat spacetimes near null infinity in accordance with Friedrich's work on the hyperboloidal initial value problem for the Einstein equations \cite{friedrich1983cauchy}. Hyperboloidal surfaces have asymptotically hyperbolic geometry in asymptotically flat spacetimes. In asymptotically de Sitter spacetimes, the term hyperboloidal is occasionally used for null-transverse hypersurfaces across the cosmological horizon \cite{mavrogiannis2023morawetz, sarkar2023perturbing} but these surfaces do not have asymptotically hyperbolic geometry; they are horizon-penetrating. The term null-transverse covers these separate cases.

In this section, we review the construction of bridge foliations in Schwarzschild--de Sitter spacetime. Bridge foliations connect the black-hole horizon with the far away observer modeled at the future cosmological horizon. We also discuss suitable coordinates for the flat limit in which the cosmological length scale goes to infinity. 

\subsubsection{The metric and the tortoise coordinate}
\begin{figure}[h]%
    \centering
    \includegraphics[width=0.6\textwidth]{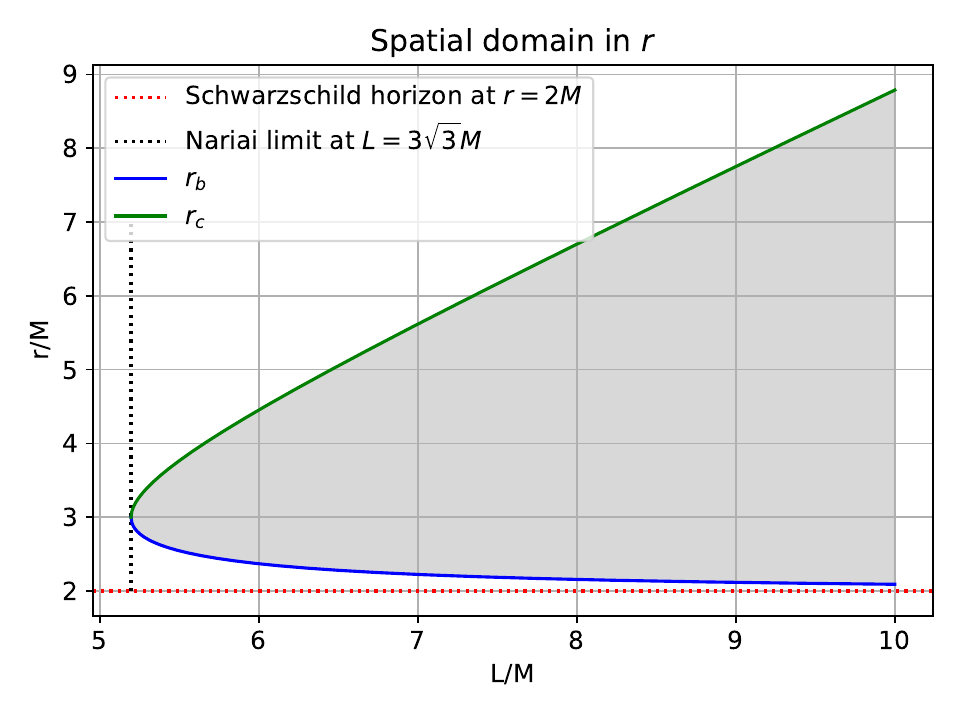}
    \caption{The spatial domain in $r$ of Schwarzschild--de Sitter spacetime grows linearly with $L$ in the flat limit, $L\to\infty$.}\label{fig:domain}
    \end{figure}

In coordinates adapted to the time and spherical symmetries, the Schwarzschild--de Sitter metric has the form \eqref{eq:metric} with
\[ f(r) = 1 - \frac{2M}{r} - \frac{r^2}{L^2},\]
where, $M$ is the black-hole mass and $L$ is the cosmological length scale related to the cosmological constant $\Lambda$ via $\Lambda = 3/L^2$. The horizons of this geometry are given by the roots of $f(r)$, solving the cubic equation, $r^3 - L^2 r + 2 M L^2 = 0$. For $M=0$, we recover the cosmological horizon of de Sitter space discussed in Sec.~\ref{sec:desitter}. For $M\in [0, L/(3\sqrt{3})]$, we get two positive real solutions corresponding to the black hole horizon, $r_b$, and the cosmological horizon, $r_c$. The negative real solution is considered unphysical and reads $r_u = -(r_b+r_c)$. The explicit form of the positive real solutions can be written as
\[ r_b = \frac{2L}{\sqrt{3}} \sin \left[\frac{1}{3} \arcsin \frac{3\sqrt{3}M}{L}\right] , \qquad r_c = \frac{2L}{\sqrt{3}} \sin \left[\frac{1}{3} \left(\pi - \arcsin \frac{3\sqrt{3}M}{L}\right) \right]. \]
We are primarily interested in the region $[r_b,r_c]$, between the future black-hole horizon $\mathcal{H}^+$ and the cosmological horizon $\mathcal{H}_c^+$. As the observed cosmological constant of our Universe is very small, we need to consider the flat limit. In the flat limit as $L\to\infty$, we recover the Schwarzschild horizon at $r_b \to 2M$ while the cosmological horizon is pushed to infinity linearly as $r_c \sim L \to \infty$, giving us an unbounded domain. The spatial domain in $r$ as a function of $L$ is depicted in Fig.~\ref{fig:domain}.

For the two-parameter Schwarzschild--de Sitter metric, we can either use $M$ and $L$ as free parameters or $r_b$ and $r_c$. The relationship between these parametrizations is
\[ L^2 = r_b^2 + r_b\, r_c + r_c^2 = r_i^2-r_j r_k, \qquad 2 M L^2 = r_b r_c (r_b + r_c)= -r_i r_j r_k,\]
where the indices $i,j,k$ stand for any distinct combination of the indices $b,c,u$.

We can write the function $f$ in terms of its roots as
\[ 
 f = \frac{1}{L^2 r} (r-r_b)(r_c-r)(r-r_u)    
\]
To write down the expression for the tortoise coordinate, it is helpful to define the surface gravities associated with each root $r_i$ of $f(r)$ as 
\[\kappa_i = \frac{1}{2} \left|\frac{df}{dr}\right|_{r=r_i} = \frac{1}{2} \left|\frac{1}{r_i} - \frac{3 r_i}{L^2}\right|.\]
The flat limit $L\to \infty$ recovers the Schwarzschild surface gravity at the event horizon, as expected. The surface gravities are commonly written out as
\[
\kappa_b = \frac{\left(r_c - r_b \right) \left(r_b - r_u \right)}{2\, L^2\, r_b}, \quad
\kappa_c = \frac{\left(r_c - r_b \right) \left(r_c - r_u \right)}{2\, L^2\, r_c}, \quad
\kappa_u =\frac{\left(r_b - r_u \right) \left(r_c - r_u \right)}{2\, L^2\, (-r_u)}\, .
\]
We picked the positive sign for the surface gravity of the cosmological horizon. We write
\be\label{eqn:sds-tortoise} \frac{1}{f} = \frac{1}{2\kappa_b (r-r_b)} + \frac{1}{2\kappa_c(r_c-r)} + \frac{1}{2\kappa_u (r-r_u)}. \ee
The tortoise coordinate reads
\[ r_*(r)= \int \frac{dr}{f(r)} = \frac{1}{2 \kappa_b} \ln \left|\frac{r}{r_b} - 1\right| - \frac{1}{2 \kappa_c} \ln\left| 1 - \frac{r}{r_c}\right| + \frac{1}{2 \kappa_u} \ln \left|1-\frac{r}{r_u} \right| \, .\]
The integration constant in the above integral is fixed to satisfy $r_\ast(0) = 0$.

\subsubsection{Examples}

We can read off the required behavior of a bridge foliation near the horizons similarly as in  Schwarzschild spacetime by considering the tortoise coordinate. Near the event horizon at $r=r_b$, the leading-order behavior for a future-directed, horizon-penetrating foliation is
\be\label{eqn:ssds_horizon} h \sim +r_\ast \sim \frac{1}{2 \kappa_b} \ln \left|\frac{r}{r_b} - 1\right|, \quad \mathrm{as} \quad r\to r_b.\ee
Near the cosmological horizon at $r=r_c$, the leading-order behavior for a future-directed, horizon-penetrating foliation is
\be\label{eqn:ssds_cosmos} h \sim  - r_\ast \sim \frac{1}{2 \kappa_c} \ln \left|1-\frac{r}{r_c}\right|, \quad \mathrm{as} \quad r\to r_c. \ee

\begin{figure}[h]%
    \centering
    \includegraphics[width=0.49\textwidth]{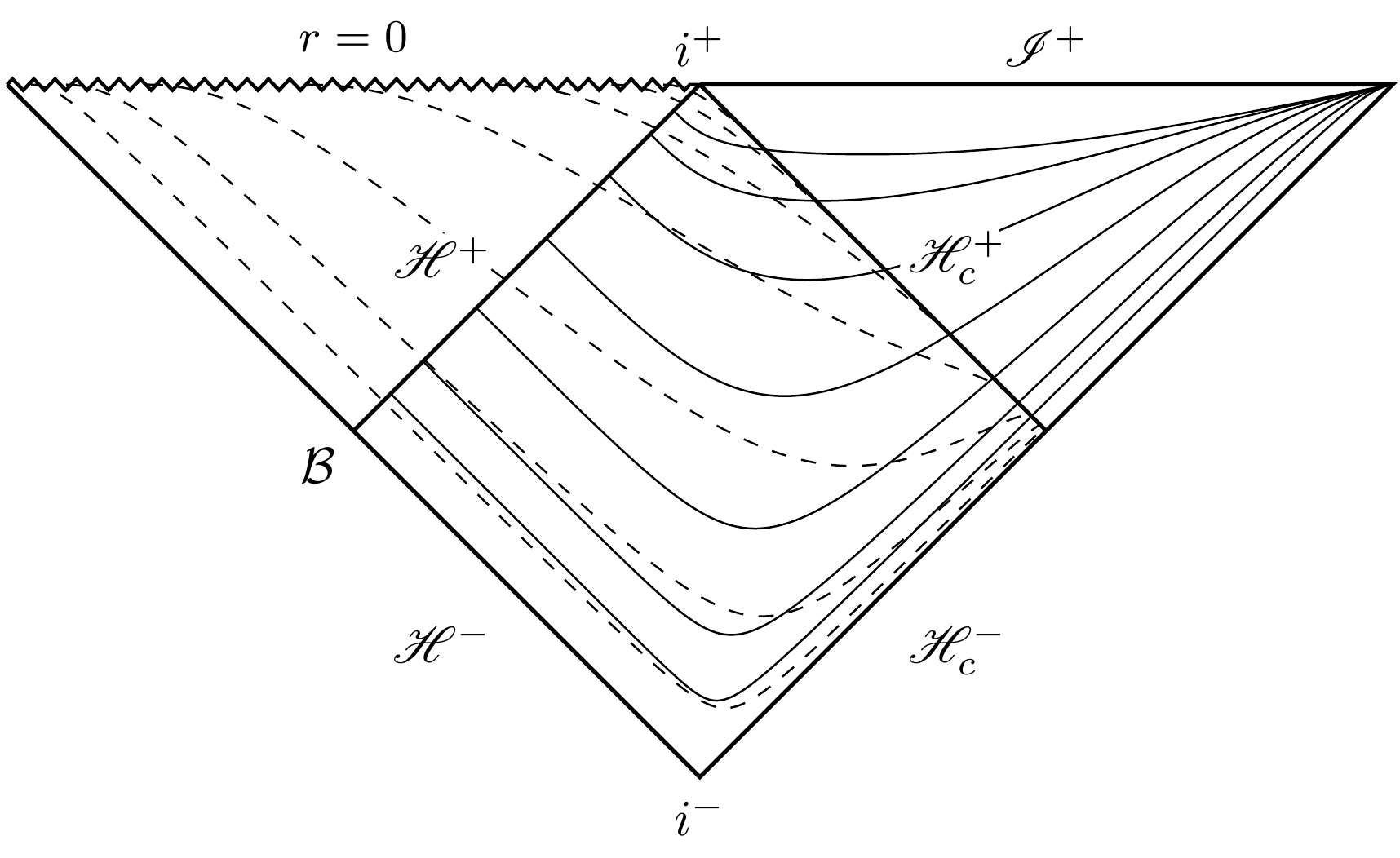}\hfill
    \includegraphics[width=0.49\textwidth]{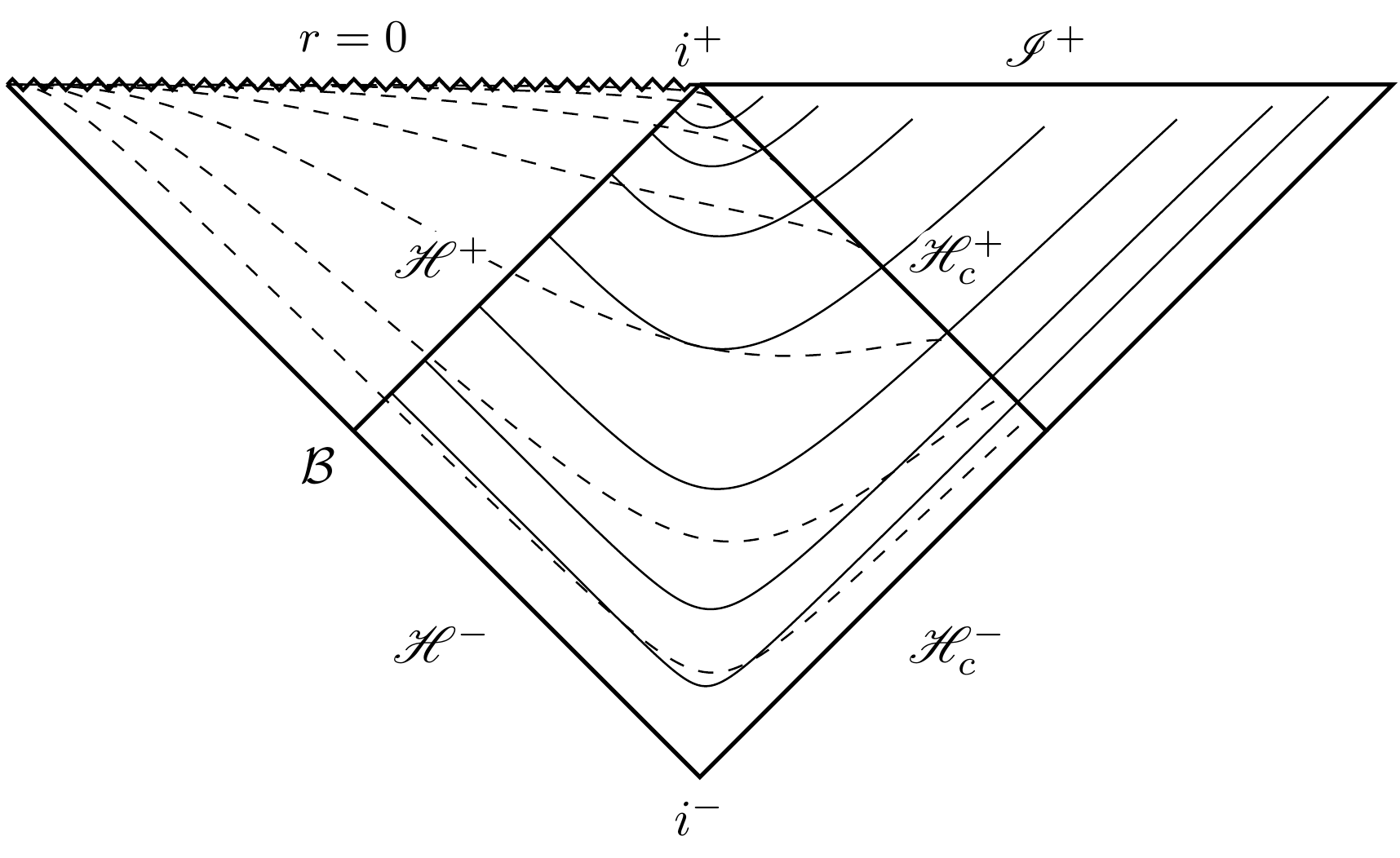}
    \caption{The conformal diagrams for Schwarzschild--de Sitter spacetime depict the bridge foliations \eqref{eqn:minimal_ssds} on the left and \eqref{eqn:rodrigo_minimal_ssds} on the right. For each case, we depict two sets of foliations corresponding to the two surface gravities required for the construction of the conformal diagram. The dashed lines are constructed using $\kappa_b$ and are regular across the black-hole horizon; the solid lines are constructed using $\kappa_c$ and are regular across the cosmological horizon.
    } \label{fig:sds_minimal}
\end{figure}

\begin{example}[Minimum height function]
The minimum necessary terms for regularizing the singular time $t$ across both null horizons is the combination of \eqref{eqn:ssds_horizon} and \eqref{eqn:ssds_cosmos}:
\be\label{eqn:minimal_ssds}
h = \frac{1}{2 \kappa_b} \ln \left|\frac{r}{r_b} - 1\right| + \frac{1}{2 \kappa_c} \ln \left|1-\frac{r}{r_c}\right|.
\ee
This construction is the same as for the minimal gauge in  Schwarzschild spacetime \eqref{eqn:minimal_gauge} in the sense that we are only considering the leading order behavior. The foliation is null-transverse on both horizons, as shown on the left panel of Fig.~\ref{fig:sds_minimal}. In the case of Schwarzschild--de Sitter, this gauge is not the same as the minimal gauge because of differences in the construction algorithm \cite{PanossoMacedo:2023qzp}.
\end{example}

\begin{example}[Minimal gauge]
The minimal gauge adds more terms to the height function, which ensures the regularity of the bridge foliation at the flat limit. We have \cite{PanossoMacedo:2023qzp},
\be\label{eqn:rodrigo_minimal_ssds}
h = \frac{1}{2 \kappa_b} \ln \left|\frac{1}{r_b} - \frac{1}{r}\right| + \frac{1}{2 \kappa_c} \ln \left|\frac{1}{r}-\frac{1}{r_c}\right| - \frac{1}{2 \kappa_u} \ln \left|\frac{1}{r}-\frac{1}{r_u}\right|.
\ee
\end{example}
These two gauges have a different asymptotics for $r\to\infty$. As depicted on the right panel of Fig.~\ref{fig:sds_minimal}, the minimal gauge opens up the slices near $\scri^+$, which allows it to have a flat limit that approaches the minimal gauge in Schwarzschild spacetime.

\begin{example}[Linear regularization]
We can also construct a regularization of both horizons using a linear relationship for the boost function:
\be\label{eq:linear} f H = - 1 + 2 \frac{r_c-r}{r_c-r_b}.\ee
The turning point is in the middle of the domain at $r_t = r_b + \frac{r_c-r_b}{2}$. The height function reads
\[ h(r) = \frac{1}{2\kappa_b} \ln |r-r_b| + \frac{1}{2\kappa_c} \ln |r_c-r| - \frac{3}{2\kappa_u} \frac{r_b+r_c}{r_b-r_c} \ln |r-r_u|. \]
The linear boost in \eqref{eq:linear} does not have a hyperboloidal flat limit as $r_c\to\infty$ but this can be achieved by a simple modification,
\[ f H = - 1 + 2  \frac{r_c-r}{r_c-r_b} \frac{r_b^2}{r^2}. \]
With this modification, the flat limit $r_c\to\infty$ gives the minimal gauge in Schwarzschild spacetime \eqref{eqn:minimal_gauge}. In the conformal diagram, the modification opens up the time slices beyond the cosmological horizon similar to the right panel of Fig.~\ref{fig:sds_minimal}.
\end{example}

\begin{example}[Mavrogiannis] Mavrogiannis constructed the bridge foliation \eqref{eqn:mavro} originally for the analysis of wave equations in Schwarzschild--de Sitter spacetime \cite{mavrogiannis_quasilinear_2021, mavrogiannis2023morawetz}. The boost function reads
\be fH = \frac{1-\frac{3M}{r}}{\sqrt{1-9M^2 \Lambda}}\sqrt{1+\frac{6M}{r}}. \ee
The boost function gives the hyperboloidal bridge \eqref{eqn:mavro} at its flat limit, $\Lambda=0$. 
\end{example}

\begin{example}[Slow-roll coordinates] As a final example, we list a bridge foliation that was developed to study the effects of horizons on quantum fields \cite{chadburn_time_2014, gregory_black_2017, gregory_evolving_2018, anderson_linear_2022, anderson_horizons_2022}. The slow-roll coordinates are defined through the boost function via
\[ 
    f H = -\gamma r + \frac{\beta}{r^2}, \qquad \textrm{with} \qquad \gamma = \frac{r_c^2+r_b^2}{r_c^3-r_b^3}, \quad \beta = \frac{r_c^2 r_b^2 (r_c+r_b)}{r_c^3-r_b^3}.    
\]
\end{example}
\section{Discussion and Outlook}\label{sec5}


The main claim of this paper may be summed up with an almost trivial statement: we should use regular coordinates across all null horizons. While the demand for regularity of coordinates is neither new nor controversial in general relativity, its application specifically to conformally extended manifolds has only recently gained attention. The intersection of standard time slices at spatial infinity has been recognized after Penrose's work in the 1960s. Null coordinates are regular across null infinity and have been used extensively. But it took almost half a century for the first applications of regular time functions across null infinity to appear in the literature. As we have shown, these developments are now playing a significant role in black-hole physics.

Of course, many coordinate choices are possible, and no single system is universally optimal. Yet, when studying dynamical fields across horizons—such as gravitational waves or quasinormal modes—the regularity of coordinates in relevant regions becomes essential. This paper reviews and advocates a specific \textit{class} of time slicing in black-hole perturbation theory: stationary bridge foliations. The particular choice of the bridge foliation may be a matter of convenience, elegance, or numerical efficiency. Future studies will investigate the optimal choices of parameters and foliations for given physical scenarios. However, the regularity of the coordinate system across the horizons is essential.

In static spacetimes, the use of regular time coordinates has an interesting consequence: it breaks time-reflection symmetry.  Far from contradicting general covariance, this reflects the broader principle that gauge freedom is constrained by boundaries. Once a class of observers aligned with the boundary is selected, the gauge freedom is naturally restricted [cite].

In static spacetimes, the use of regular time coordinates has the side effect of breaking time-reflection symmetry. Intuitively, this is related to the impossibility of synchronizing clocks across horizons via two-way communication. It does not contradict the principle of general covariance; rather, it reflects the broader fact that gauge freedom is restricted by the presence of boundaries. Once a particular class of observers that agrees with the position of the boundary is selected, the gauge freedom becomes broken \cite{compère2019advancedlecturesgeneralrelativity, fiorucci2021leakycovariantphasespaces}. 

The insights from this work can be extended beyond stationary, spherically symmetric spacetimes. The essential features of bridge foliations rely only on asymptotic conditions, making them applicable to more general black-hole solutions. However, the dynamical case—requiring solutions to the nonlinear Einstein equations—presents significant challenges. While the excision method across black-hole horizons is well-established, the analogous problems at null infinity and cosmological horizons remain open. The pioneering work of Friedrich in 1983 laid the foundation for analyzing the hyperboloidal initial value problem \cite{friedrich1983cauchy}. Significant effort has been made to apply some of his ideas in numerical relativity \cite{frauendiener2004conformal}. The numerical solution of Einstein equations for hyperboloidal data is currently an active area of research with promising developments in both theoretical and computational approaches \cite{gautam2021summation, peterson20233d, vano2023spherically, peterson2024spherical, vano2024height}.

\section*{Acknowledgements}
I thank David Hilditch, Rodrigo Panosso Macedo, and Alex Vañó-Viñuales for discussions and comments on the manuscript. This material is supported by the National Science Foundation under Grant No.~2309084.

\bibliography{refs.bib}
\end{document}